\documentclass[prd,twocolumn,showpacs,preprintnumbers,nofootinbib,amsmath,amssymb,superscriptaddress]{revtex4}

\usepackage{graphicx}
\usepackage{dcolumn,color}
\usepackage{bm,hyperref}
\usepackage{slashed}
\usepackage{braket}
\usepackage{float}

\usepackage[utf8]{inputenc}

\begin{document}

\title{Precision constraints on radiative neutrino decay with CMB
spectral distortion}

\author{Jelle L. Aalberts}
\affiliation{Institute for Theoretical Physics, 
University of Amsterdam, 1098 XH Amsterdam, The Netherlands}
\author{Shin'ichiro Ando}
\affiliation{GRAPPA Institute, University of Amsterdam, 1098 XH
Amsterdam, The Netherlands}
\affiliation{Institute for Theoretical Physics, 
University of Amsterdam, 1098 XH
Amsterdam, The Netherlands}
\affiliation{Kavli Institute for the Physics and Mathematics of the
Universe (Kavli IPMU, WPI), Todai Institutes for Advanced Study,
University of Tokyo, Kashiwa, Chiba 277-8583, Japan}
\author{Wouter M. Borg}
\affiliation{Institute for Theoretical Physics, 
University of Amsterdam, 1098 XH Amsterdam, The Netherlands}
\author{Edwin Broeils}
\affiliation{Institute for Theoretical Physics, 
University of Amsterdam, 1098 XH Amsterdam, The Netherlands}
\author{Jennypher Broeils}
\affiliation{Institute for Theoretical Physics, 
University of Amsterdam, 1098 XH Amsterdam, The Netherlands}
\author{Stephen Broeils}
\affiliation{Institute for Theoretical Physics, 
University of Amsterdam, 1098 XH Amsterdam, The Netherlands}
\author{Bradley J. Kavanagh}
\affiliation{GRAPPA Institute, University of Amsterdam, 1098 XH
Amsterdam, The Netherlands}
\affiliation{Institute for Theoretical Physics, 
University of Amsterdam, 1098 XH Amsterdam, The Netherlands}
\author{Gijs Leguijt}
\affiliation{Institute for Theoretical Physics, 
University of Amsterdam, 1098 XH Amsterdam, The Netherlands}
\author{Marnix Reemst}
\affiliation{Institute for Theoretical Physics, 
University of Amsterdam, 1098 XH Amsterdam, The Netherlands}
\author{Dylan R. van Arneman}
\affiliation{Institute for Theoretical Physics, 
University of Amsterdam, 1098 XH Amsterdam, The Netherlands}
\author{Hoang Vu}
\affiliation{Institute for Theoretical Physics, 
University of Amsterdam, 1098 XH Amsterdam, The Netherlands}

\date{\today}

\begin{abstract}
We investigate the radiative decay of the cosmic neutrino background,
 and its impact on the spectrum of the cosmic microwave background (CMB) that is
 known to be a nearly perfect black body.
We derive {\it exact} formulae for the decay of a heavier neutrino into
 a lighter neutrino and a photon, $\nu_j \to \nu_i + \gamma$, and of
 absorption as its inverse, $\nu_i + \gamma \to \nu_j$, by accounting for
 the precise form of the neutrino momentum distribution.
Our calculations show that if the neutrinos are heavier than $\mathcal
 O(0.1)$~eV, the exact formulae give results that differ by $\sim$50\%,
 compared with approximate ones where neutrinos are assumed to be at
 rest.
We also find that spectral distortion due to absorption is more
 important for heavy neutrino masses (by a factor of $\sim$10 going from
 a neutrino mass of 0.01~eV to 0.1~eV).
By analyzing the CMB spectral data measured with COBE-FIRAS, we obtain
 lower limits on the neutrino lifetime of  $\tau_{12} \gtrsim 4 \times
 10^{21}$~s (95\% C.L.) for the smaller mass splitting and $\tau_{13}
 \sim \tau_{23} \gtrsim 10^{19}$~s for the larger mass splitting.
These represent up to one order of magnitude improvement over previous
 CMB constraints.
With future CMB experiments such as PIXIE, these limits will improve by
 roughly 4 orders of magnitude.
This translates to a projected upper limit on the neutrino magnetic
 moment (for certain neutrino masses and decay modes) of $\mu_\nu < 3
 \times 10^{-11}\, \mu_B$, where $\mu_B$ is the Bohr magneton.
Such constraints would make future precision CMB measurements
 competitive with lab-based constraints on neutrino magnetic moments.
\end{abstract}


\maketitle

\section{Introduction}
\label{sec:introduction}

In the last few decades, many experiments have demonstrated that
neutrinos show properties beyond the Standard Model of particle
physics.
They have nonzero masses and show flavor mixings as revealed by
measurements of neutrino oscillations using solar, atmospheric, reactor,
and accelerator neutrinos (see Refs.~\cite{PDG,Giganti:2017fhf} for a
review).
There are, however, a number of important issues remaining:
What is the neutrino mass hierarchy~\cite{Qian:2015waa}?
What is the CP violating phase in the lepton
sector~\cite{Hagedorn:2017wjy}? 
How weakly do neutrinos interact with
photons~\cite{Levine1967,Cung1975}?
Do neutrinos decay, either radiatively or
non-radiatively~\cite{Pal:1981rm}?

Even the weak interaction predicts interactions between the neutrino and
photon through a non-zero magnetic moment induced via loop corrections
of gauge boson, although its value is expected to be small \cite{Cung1975}. 
We denote by $\mu_{ij}$ the magnetic moment between neutrino mass
eigenstates $i$ and $j$, with off-diagonal elements $(i\neq j)$
representing the transition magnetic moments in radiative decay.
For massive Dirac neutrinos, the value of the diagonal magnetic moment
induced by loops of gauge bosons is given
by~\cite{Fujikawa1980}
\begin{equation}
\mu^{D}_{ii} =
\frac{3eG_{F}m_{i}}{8\sqrt{2}\pi^{2}}\approx
3.2\times10^{-19}\left(\frac{m_{i}}{\mathrm{eV}}\right)\,\mu_{B},
\end{equation}
where $\mu_B$ is the Bohr magneton, while for the Dirac off-diagonal
elements one finds a value roughly $10^{-4}$ times
smaller~\cite{Glashow1970}.
For Majorana neutrinos, one finds the magnetic moment suppressed by the
ratio of the lepton and gauge boson masses:
\begin{equation}
\mu^{M}_{ij} =
\frac{3eG_{F}m_{i}}{16\sqrt{2}\pi^{2}}\left(1+\frac{m_{j}}{m_{i}}\right)\sum_{l=e,\mu
, \tau}
{\rm Im}\left[U_{lk}U_{lj}^{*}\right]\left(\frac{m_{l}}{m_{W}}\right)^{2} \,.
\end{equation}

The current upper bound on the neutrino magnetic moment from
electron-neutrino scattering experiments is $\mu_{\nu}<2.8 \times
10^{-11}\mu_{B}$~\cite{Beda2013,Borexino:2017fbd}. 
The strongest astrophysical constraints place the bound at
$\mu_{\nu}<2.2 \times
10^{-12}\mu_{B}$~\cite{Raffelt1990,Raffelt:1999gv,Arceo-Diaz:2015pva},
well above the value expected from weak interactions alone (see
Refs.~\cite{Studenikin:2016ykv,Studenikin:2018vnp} for a thorough
review).
However, new physics contributions could enhance the predicted magnetic
moment~\cite{Shrock:1974nd,Lee:1977tib,Shrock:1982sc,Georgi:1990se,Davidson:2005cs,Bell:2005kz,Bell:2006wi,Frere:2015pma}.
In particular, Ref.~\cite{Lindner:2017uvt} proposed a model with SU(2)
horizontal symmetry, allowing Majorana transition neutrino magnetic
moments of order $10^{-12}\,\mu_B$ while protecting the small mass of
the neutrinos.

Since the neutrinos have masses that are proven to differ for different
mass eigenstates, nonzero magnetic moments will induce radiative
neutrino decay,
\begin{equation}
 \nu_j \to \nu_i + \gamma,
\end{equation}
where $m_j > m_i$.
With enhanced magnetic moments, radiative neutrino decays induced
by this interaction may be relevant for astrophysical systems, providing
a probe of new physics in the neutrino sector.
Since the mass squared differences have been precisely measured with
oscillation experiments and the absolute neutrino masses have been
constrained to be below $\mathcal
O(1)$~eV~\cite{Ade:2015xua,KamLAND-Zen:2016pfg}, Ref.~\cite{Mirizzi2007}
pointed out that photons emitted via neutrino decay will disturb the
nearly perfect black-body spectrum of the cosmic microwave background
(CMB).
By comparing with the CMB spectral data obtained with the Far Infrared
Absolute Spectrophotometer (FIRAS) onboard the Cosmic Background
Explorer (COBE)~\cite{Fixsen:1996nj, 2002ApJ...581..817F},
Ref.~\cite{Mirizzi2007} obtained constraints on decay rate as $\Gamma <
2\times 10^{-19}$--$5\times 10^{-20}$ s$^{-1}$.
This corresponds to $\mu_\nu \alt 10^{-8} \mu_B$, still much weaker than
other astrophysical or lab-based constraints.

In this paper, we improve the work of Ref.~\cite{Mirizzi2007} in several
aspects.
First, we argue that, if the neutrino can decay radiatively ($\nu_j \to
\nu_i + \gamma$), then it is also possible that a CMB photon is absorbed
by a lighter mass eigenstate of the cosmic neutrino background $\nu_i$:
\begin{equation}
\nu_i + \gamma_{\rm CMB} \to \nu_j,
\end{equation}
to create a heavier state $\nu_j$.
The cross section of this resonance process is given by
\begin{equation}
\sigma(E) = \frac{\pi^2}{k^2}\Gamma \delta (E-m_j),
\label{eq:absorption cross section}
\end{equation}
where $E$ is the center-of-mass energy of the initial state, and $k$ is
the momentum of the center-of-mass frame~\cite{PDG}.

Second, in contrast to approximate formulae that were derived and
adopted in the literature~\cite{Masso:1999wj, Mirizzi2007}, where
neutrinos were assumed to be at rest, we derive exact formulae by taking
the neutrinos' thermal momentum distribution into account.
We also include the effects of stimulated emissions for decay and
Pauli blocking for both the decay and absorption.
We show that all these effects can be of considerable importance in
calculating the CMB spectral distortion.
Therefore, neglecting these will cause a theoretical bias in the estimated
lower limits on the decay lifetime.

Third, we will make projections for planned future CMB experiments. 
We will primarily focus on the Primordial Inflation Explorer
(PIXIE)~\cite{2011JCAP...07..025K}, which is a proposed mission to
measure the CMB intensity with a higher sensitivity and wider frequency
range than COBE-FIRAS. 
PIXIE is expected to have a sensitivity of 5~Jy/sr~\cite{Chluba1}, as
opposed to the COBE-FIRAS sensitivity, which is of the order of
$10^{4}$~Jy/sr~\cite{Fixsen:1996nj}. 
We show how this improved sensitivity affects constraints on the
neutrino lifetime and magnetic moment. 
We thus motivate future CMB experiments such as PIXIE (and the
further-future PRISM experiment~\cite{PRISMcollab}), as probes of New
Physics in the neutrino sector.
Finally, we note that, when obtaining the current FIRAS bounds and PIXIE
sensitivities, we take into account correlations between different
components of the spectral distortion such as the chemical potential as
well as the residual Galactic emission.

The paper is organized as follows.
In Sec.~\ref{sec:formulation}, we present formulae for computing the
intensities of microwave photons from both decay and absorption by
cosmic neutrinos.
The theoretical results are then compared with the CMB spectral data
measured with COBE-FIRAS, and we calculate the lower limits on the decay
lifetime and the upper limits on the neutrino magnetic moments for
various modes and the different mass hierarchy scenarios in
Sec.~\ref{sec:analysis}.
We then discuss the potential sensitivity of future CMB experiments to the
neutrino radiative decay in Sec.~\ref{sec:future} and conclude the paper
in Sec.~\ref{sec:conclusions}.


\section{Decay and absorption intensities}
\label{sec:formulation}

In Sec.~\ref{sec:approximate formulae}, we first derive formulae for
photon intensities from both the decay of heavier neutrinos and
absorption of the CMB photons by the lighter neutrinos, by assuming
that the neutrinos are at rest --- a reasonable approximation when the
neutrino mass is much larger than the temperature of the CMB and
neutrino background, on the order of $10^{-3}$~eV.
In Sec.~\ref{sec:exact formulae}, we show exact formulae for the decay
and absorption intensities although much of the derivation is later
summarized in Appendix~\ref{sec:appendix A}.
In Sec.~\ref{sec:theory results}, we show numerical results for decay
and absorption intensities, illustrating their dependence on the
lightest neutrino mass, decay mode, and mass hierarchy.

\subsection{Approximate formulation}
\label{sec:approximate formulae}

If the neutrino can be considered at rest, according to kinematics, the
energy of the absorbed CMB photon $\epsilon_\gamma$ (in the observer
frame) is related to the neutrino masses via
\begin{equation}
(1+z_a) \epsilon_\gamma = \frac{\Delta m_{ij}^2}{2m_i},
\label{eq:absorption energy}
\end{equation}
where we have defined $m_i < m_j$, $\Delta m_{ij}^2 \equiv m_j^2 - m_i^2$,
and $z_a$ is the redshift when absorption occurs.
In contrast, radiative decay follows slightly different kinematics:
\begin{equation}
(1+z_d) \epsilon_\gamma = \frac{\Delta m_{ij}^2}{2m_j},
\label{eq:decay energy}
\end{equation}
with redshift $z_d$ when the decay occurs. Setting $z_{a,d} = 0$ gives the maximum photon energy at which the effects of absorption or decay can be observed for given $m_i$, $m_j$.

Equation~(\ref{eq:absorption cross section}) shows the absorption cross
section in terms of quantities in the center-of-mass frame.
It is, however, more useful to use quantities in the observer frame
where the lighter neutrino $\nu_i$ is at rest and $\epsilon_\gamma$ is
in the microwave frequency range.
Then, Eq.~(\ref{eq:absorption cross section}) can be rewritten as
\begin{eqnarray}
 \sigma([1+z]\epsilon_\gamma) &=& \frac{\pi^2}{k^2}
  \Gamma   \frac{m_j}{m_i}
  \delta \left((1+z)\epsilon_\gamma - \frac{\Delta
 m_{ij}^2}{2m_i}\right)\,.
 \end{eqnarray}
The center-of-mass momentum $k$ for the absorption,
$\nu_i+\gamma\to\nu_j$, satisfies
\begin{equation}
 \sqrt{k^2+m_i^2} + k = m_j\,,
\end{equation}
according to energy conservation.
From this, we have
\begin{equation}
 k = \frac{\Delta m_{\ij}^2}{2m_j}\,,
\end{equation}
and thus
\begin{eqnarray}
 \sigma([1+z]\epsilon_\gamma)  &=& \frac{4\pi^2m_j^3\Gamma}{(\Delta m_{ij}^2)^2 m_i\epsilon_\gamma}
 \delta (z - z_a)\,.
\end{eqnarray}

We then consider the {\it effective} CMB intensity due to decay and
absorption, $I_{\rm dec}$ and $I_{\rm abs}$ as a function of CMB
energy $\epsilon_\gamma$.
For a given energy $\epsilon_\gamma$, there is a corresponding redshift
through Eqs.~(\ref{eq:absorption energy}) and (\ref{eq:decay energy}),
where absorption and decay is allowed, respectively.
These intensities can be written as a cosmological line-of-sight
integral of the emissivity~\cite{Peacock:1999ye}:
\begin{eqnarray}
I_{\rm dec}(\epsilon_\gamma) &=& \frac{1}{4\pi} \int dz \frac{P_{\rm
 dec} ([1+z]\epsilon_\gamma, z)}{H(z)(1+z)^4}\,,
 \label{eq:intensity decay}\\
I_{\rm abs}(\epsilon_\gamma) &=& \frac{1}{4\pi} \int dz \frac{P_{\rm
 abs}  ([1+z]\epsilon_\gamma, z)}{H(z)(1+z)^4}\,,
 \label{eq:intensity absorption}
\end{eqnarray}
where $P_{\rm dec}$ or $P_{\rm abs}$ is the volume emissivity (energy of
photons emitted per unit volume, per unit time, and per unit energy
range), $H(z) = H_0 \sqrt{\Omega_m(1+z)^3+\Omega_\Lambda}$, $H_0 =
67.8$~km~s$^{-1}$~Mpc$^{-1}$ is the Hubble constant, $\Omega_m = 0.308$,
and $\Omega_\Lambda = 0.692$~\cite{Ade:2015xua}.

These emissivity functions can therefore be written as
\begin{eqnarray}
P_{\rm dec}([1+z]\epsilon_\gamma, z) &=& (1+z) \epsilon_\gamma
 n_{\nu_j}(z) \Gamma  e^{-\Gamma t(z)}
 \nonumber\\&&{}\times
 \left[1+f_{\rm CMB}(\epsilon_\gamma)\right]
 \nonumber\\&&{}\times
 \delta \left((1+z)\epsilon_\gamma - \frac{\Delta
 m_{ij}^2}{2m_j}\right)
 \nonumber\\&=&
 (1+z)n_{\nu_j}(z)\Gamma e^{-\Gamma t(z)} 
 \nonumber\\&&{}\times
 \left[1+f_{\rm CMB}(\epsilon_\gamma)\right]
 \delta (z-z_d),
 \label{eq:emissivity decay}\\
P_{\rm abs}([1+z]\epsilon_\gamma, z) &=& -(1+z) \epsilon_\gamma
 n_{\nu_i}(z) 
 \nonumber\\&&{}\times
 n_{\rm CMB}([1+z]\epsilon_\gamma,z) \sigma ([1+z]\epsilon_\gamma)
  \nonumber\\&=& 
 -\frac{2\pi^2 m_j^3 \Gamma}{\Delta m_{ij}^2m_i^2\epsilon_\gamma}
 \delta (z-z_a)
 \nonumber\\&&{}\times
 n_{\nu_i}(z) n_{\rm CMB}([1+z]\epsilon_\gamma,z),
 \label{eq:emissivity absorption}
\end{eqnarray}
where note that the sign of $P_{\rm abs}$ is negative as it gives
suppression of the total CMB intensity.
The term $\left[1+f_{\rm CMB}(\epsilon_\gamma)\right]$ represents the
stimulated emission, with $f_{\rm CMB}(\epsilon_\gamma) =
(e^{{\epsilon_\gamma}/T_{\rm CMB}}-1)^{-1}$ the occupation number
of the CMB photons and $T_{\rm CMB} = 2.725$~K the present CMB
temperature~\cite{Fixsen:1996nj, 2002ApJ...581..817F},
and $n_{\rm CMB}(\epsilon_\gamma, z)$ is the CMB number density per unit
energy range at $\epsilon_\gamma$ and $z$: i.e., $n_{\rm
CMB}([1+z]\epsilon_\gamma,z) = (1+z)^2\epsilon_\gamma^2
f_{\rm CMB}(\epsilon_\gamma)/\pi^2$.
The occupation number has no dependence on redshift, as it cancels
between the energy at $z$, $(1+z)\epsilon_\gamma$, and the CMB
temperature at $z$, $(1+z)T_{\rm CMB}$.
We note that the effect of stimulated emission has not been taken into
account in the literature~\cite{Mirizzi2007, Masso:1999wj}, although
it was acknowledged in Ref.~\cite{Masso:1999wj}.
We also assume $\Gamma t(z)\ll 1$ in the following discussions, which is
well justified when the lifetime $\tau = \Gamma^{-1}$ is much larger
than the age of the Universe as is the case here.
By using Eqs.~(\ref{eq:emissivity decay}) and (\ref{eq:emissivity
absorption}) in Eqs.~(\ref{eq:intensity decay}) and (\ref{eq:intensity
absorption}) respectively, one can predict the effect of decay and
absorption on the CMB intensity spectrum.
After the $\delta$-functions collapse the redshift integral, we obtain
the following analytic expressions:
\begin{eqnarray}
 I_{\rm dec}(\epsilon_\gamma) &=& \frac{1}{4\pi}
  \frac{n_{\nu_j}\Gamma}{H(z_d)}
  \left[1+f_{\rm CMB}(\epsilon_\gamma)\right],
  \label{eq:intensity dec approximate}
  \\
 I_{\rm abs}(\epsilon_\gamma) &=&
  -\frac{1}{4\pi}\frac{n_{\nu_i}\Gamma}{H(z_a)}\left(\frac{m_j}{m_i}\right)^3
  f_{\rm CMB}(\epsilon_\gamma),
  \label{eq:intensity abs approximate}
\end{eqnarray}
where $n_{\nu_i} = n_{\nu_j} \approx 110$~cm$^{-3}$ are the neutrino
number density of mass eigenstates $\nu_i$ and $\nu_j$ at $z = 0$.
Up to the factor for stimulated emission, Eq.~(\ref{eq:intensity
dec approximate}) agrees with the formulae adopted in
Ref.~\cite{Mirizzi2007}.


\subsection{Exact formulation}
\label{sec:exact formulae}

Thus far, we made the approximation that both $\nu_i$ and $\nu_j$ in the initial
states are at rest.
This is a very good approximation when the neutrino can be regarded as
nonrelativistic, which is valid in the case of $m_{i, j} \gg T_\nu =
(4/11)^{1/3}T_{\rm CMB} =  1.95$~K.
Otherwise, one has to take into account the momentum distribution of the
neutrinos~\cite{Wong:2011ip}:
\begin{equation}
\label{eq:momenta}
 f_{\nu}(p_\nu, z) =  \frac{1}{\exp[p_\nu/T_\nu(z)]+1},
\end{equation}
where $T_\nu(z) = (1+z) T_\nu$ is the neutrino temperature at
$z$.

A detailed derivation of the emissivity is summarized in
Appendix~\ref{sec:appendix A}, and here we show only the results:
\begin{widetext}
\begin{eqnarray}
 P_{\rm dec}([1+z]\epsilon_\gamma,z) &=& \frac{\Gamma T_\nu m_j\Delta
  m_{ij}^2}{4\pi^2} 
  \frac{1+f_{\rm  CMB}(\epsilon_\gamma)}{\epsilon_\gamma}
  U\left(\frac{m_j}{(1+z)T_\nu},\frac{\epsilon_\gamma}{T_\nu},
   \frac{2(1+z)\epsilon_\gamma m_j}{\Delta m_{ij}^2}\right),
  \label{eq:P decay exact}
  \\
 P_{\rm abs}([1+z]\epsilon_\gamma,z) &=& -\frac{2(1+z)^4}{\pi^2}
  \frac{\Gamma T_\nu^2 m_j^3}{(\Delta m_{ij}^2)^2}
  \epsilon_\gamma^2 f_{\rm CMB}(\epsilon_\gamma)
  V\left(\frac{m_i}{(1+z)T_\nu},\frac{\epsilon_\gamma}{T_\nu},
   \frac{2(1+z)\epsilon_\gamma m_i}{\Delta m_{ij}^2}\right),
  \label{eq:P abs exact}
\end{eqnarray}
\end{widetext}
for decay ($\nu_j \to\nu_i + \gamma$) and absorption ($\nu_i+\gamma\to
\nu_j$) respectively, where 
\begin{eqnarray}
 U(y,s,t) &=& \int_{\frac{y}{2}\left|t-\frac{1}{t}\right|}^\infty
  \frac{dx x}{(e^x+1)\sqrt{x^2+y^2}}
  \nonumber\\&&{}\times
 \left[1-\frac{1}{e^{W_{-}(s,t,x,y)}+1}\right],
 \\
 V(y,s,t) &=& \int_{\frac{y}{2}\left|t-\frac{1}{t}\right|}^\infty
 \frac{dx x}{e^x+1}
 \left[1-\frac{1}{e^{W_{+}(s,t,x,y)}+1}\right],
 \nonumber\\&&
 \\
 W_{\pm}(s,t,x,y) &=& \left[s^2+x^2\pm 2s
		       \left(\sqrt{x^2+y^2} -\frac{y}{t}\right)\right]^{1/2}.
 \nonumber\\&&
\end{eqnarray}
These equations are inevitably more complicated than those shown in the
previous subsection, but the most accurate.

\subsection{Results}
\label{sec:theory results}

We present numerical results for the intensity due to
absorption and decay, as well as comparing the approximate and exact
calculations. 
Because it is not yet known whether the neutrino mass eigenstates are
arranged in a normal hierarchy (NH, $m_1 < m_2 \ll m_3$) or an inverted
hierarchy (IH, $m_3 \ll m_1 < m_2$) we include both possibilities in the
calculations presented later in the paper. 
Throughout the paper, we adopt $\Delta m_{12}^2 = 7.53\times
10^{-5}$~eV$^2$ and $\Delta m_{23}^2 = 2.5\times
10^{-3}$~eV$^2$ for NH, and $\Delta m_{12}^2 = 7.53\times
10^{-5}$~eV$^2$ and $\Delta m_{31}^2 = 2.5\times
10^{-3}$~eV$^2$ for IH~\cite{PDG}.
We show only results for the NH in this section, noting that the results
for IH are similar.
The mass of the lightest neutrino mass eigenstate is $m_1$, and we
assume a reference value of $\tau = 10^{18}$~s for the neutrino
radiative decay lifetime.

\begin{figure*}[t]
    \centering
    \includegraphics[width=0.9\textwidth]{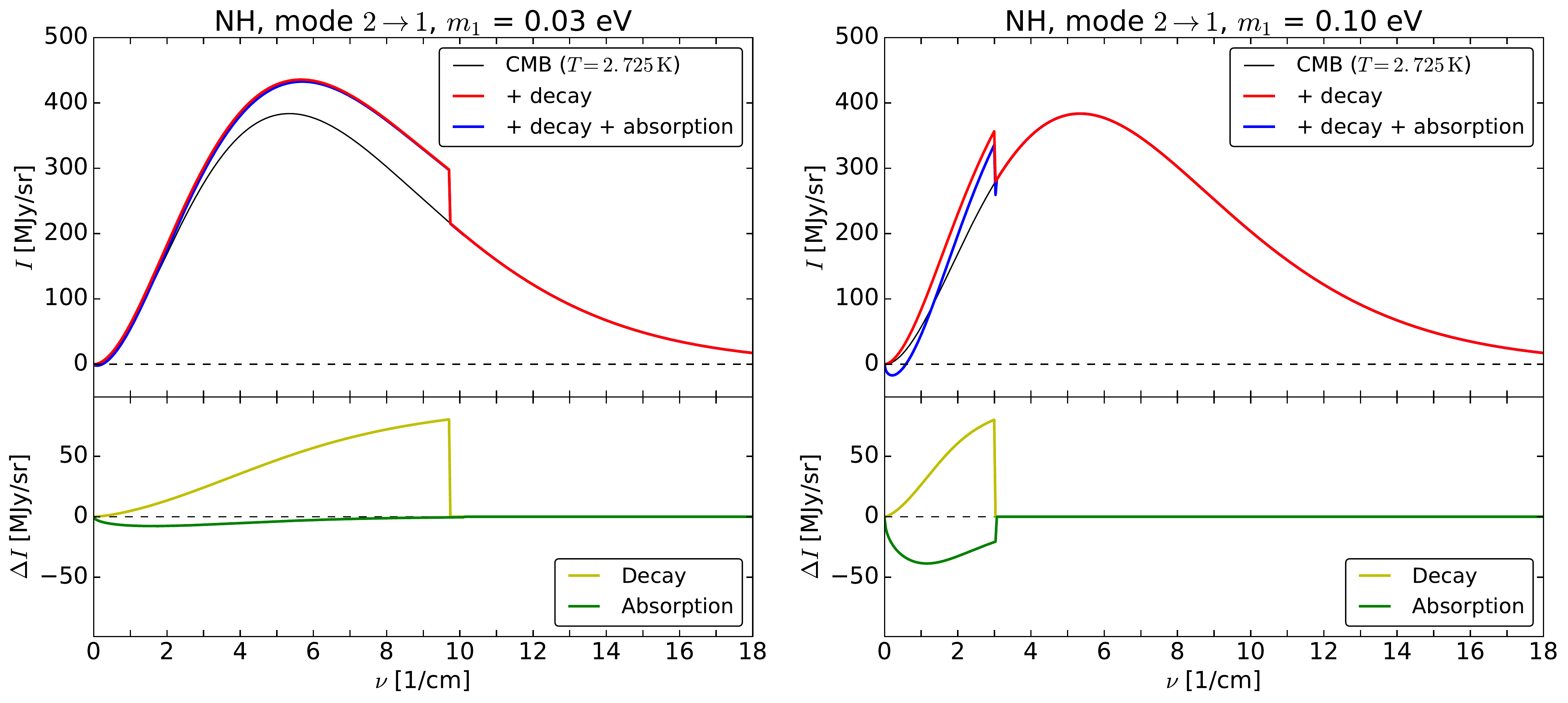}
    \caption{\textbf{Effect of neutrino decay and photon absorption on
 the CMB spectrum.} Both panels
 consider NH and $\Gamma = 10^{-18}$~s$^{-1}$ with decay/absorption
 between the lowest two mass
 eigenstates, where the lowest mass is taken to be $0.03$ eV
 (\textbf{left}) and $0.1$ eV (\textbf{right}). The black lines
 correspond to the unperturbed CMB spectrum, the red lines include the
 effect of photons from decaying neutrinos, while the blue lines
 include both decay and absorption. Here, the intensities as a function
 of frequency $\nu$ are calculated
 using the approximate formulae given in Sec.~\ref{sec:approximate
 formulae}. In the lower panel, the CMB spectrum is not included, and
 only the bare intensities from decay (yellow) and absorption (green)
 are shown.}
    \label{fig:absorp_12}
\end{figure*}

Figure~\ref{fig:absorp_12} shows the effect of decay and absorption on
the CMB spectrum, in the case of transitions between $m_1$ and $m_2$,
computed with the approximate formulae. 
We note first that the distortions to the CMB spectrum extend up to
higher frequencies for lighter neutrinos, which is simply a consequence
of kinematics [cf.~Eqs.~\eqref{eq:decay energy}~and~\eqref{eq:absorption
energy}]. 
We also note that the magnitude of the absorption depends on the masses
of the neutrinos, with heavier neutrinos leading to a larger absorption
effect. 
For a given photon energy today $\epsilon_\gamma$, as we decrease the
mass $m_i$ of the absorbing neutrino, the absorption must occur at
earlier times (larger redshift, $z_a$). 
As we increase $z_a$, the period over which absorption can take place
$\sim$$H(z_a)^{-1}$ becomes shorter, suppressing the total amount of
absorptions. Decreasing $m_i$ from 0.1 to 0.03~eV, this factor dominates 
over the $(m_j/m_i)^3$ scaling in the
absorption intensity [Eq.~\eqref{eq:intensity abs approximate}] and the
absorption effect becomes smaller
(left panel of Fig.~\ref{fig:absorp_12}).

Decreasing the lightest neutrino mass $m_i$ further, one would expect eventually that the $(m_j/m_i)^3$ scaling would dominate over the scaling with $\sim$$H(z_a)^{-1}$$\sim$$(m_i/\Delta m_{ij}^2)^{3/2}$, for high $z$. However, at high $z$ the neutrino temperature is large and the neutrino momenta, described in Eq.~\ref{eq:momenta} become relevant. The non-zero neutrino momenta act to suppress the absorption cross section [Eq.~\ref{eq:absorption cross section}] which scales as $k^{2}$, where $k$ is the centre-of-mass from momentum. As we will see in Sec.~\ref{sec:analysis}, the overall effect is that the absorption intensity flattens to a constant at small values of $m_i$. This emphasises the importance of the exact formulation -- accounting for the thermal neutrino distribution -- for the correct calculation of the absorption effect.

\begin{figure*}[t]
    \centering
    \includegraphics[width=0.9\textwidth]{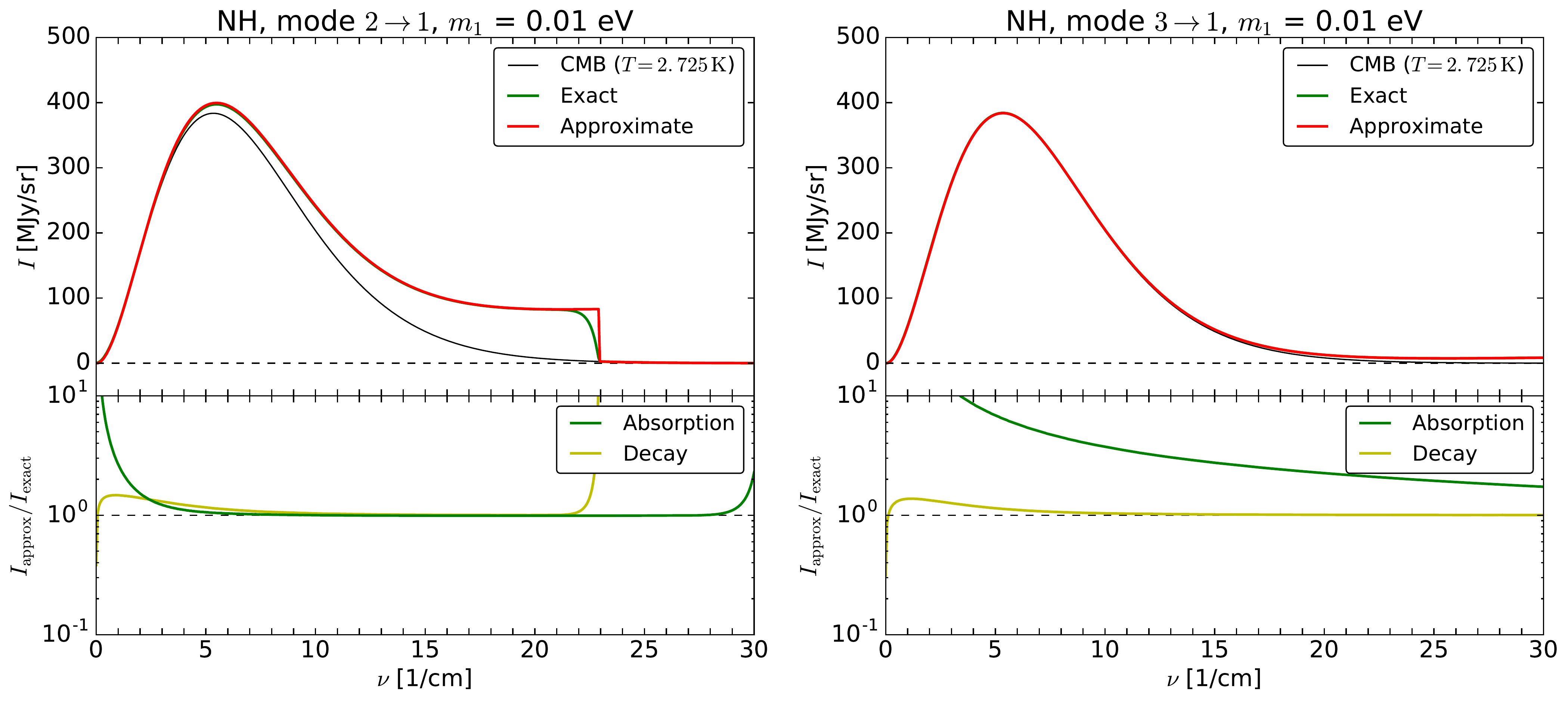}
    \caption{\textbf{Comparison between the exact and approximate
 neutrino decay and absorption intensities.} The exact results are shown
 in green and the approximate results in red, as functions of frequency
 $\nu$. Both left and right-hand panels use a mass of $0.01$ eV for the
 lightest neutrino, but they consider different modes: $m_2
 \leftrightarrow m_1$ (\textbf{left}) and $m_3 \leftrightarrow m_1$
 (\textbf{right}). The lower panels show the ratio of the approximate
 and exact intensities for both absorption and decay.}
    \label{fig:ex_vs_approx_0.01}
\end{figure*}

\begin{figure*}[t]
    \centering
    \includegraphics[width=0.9\textwidth]{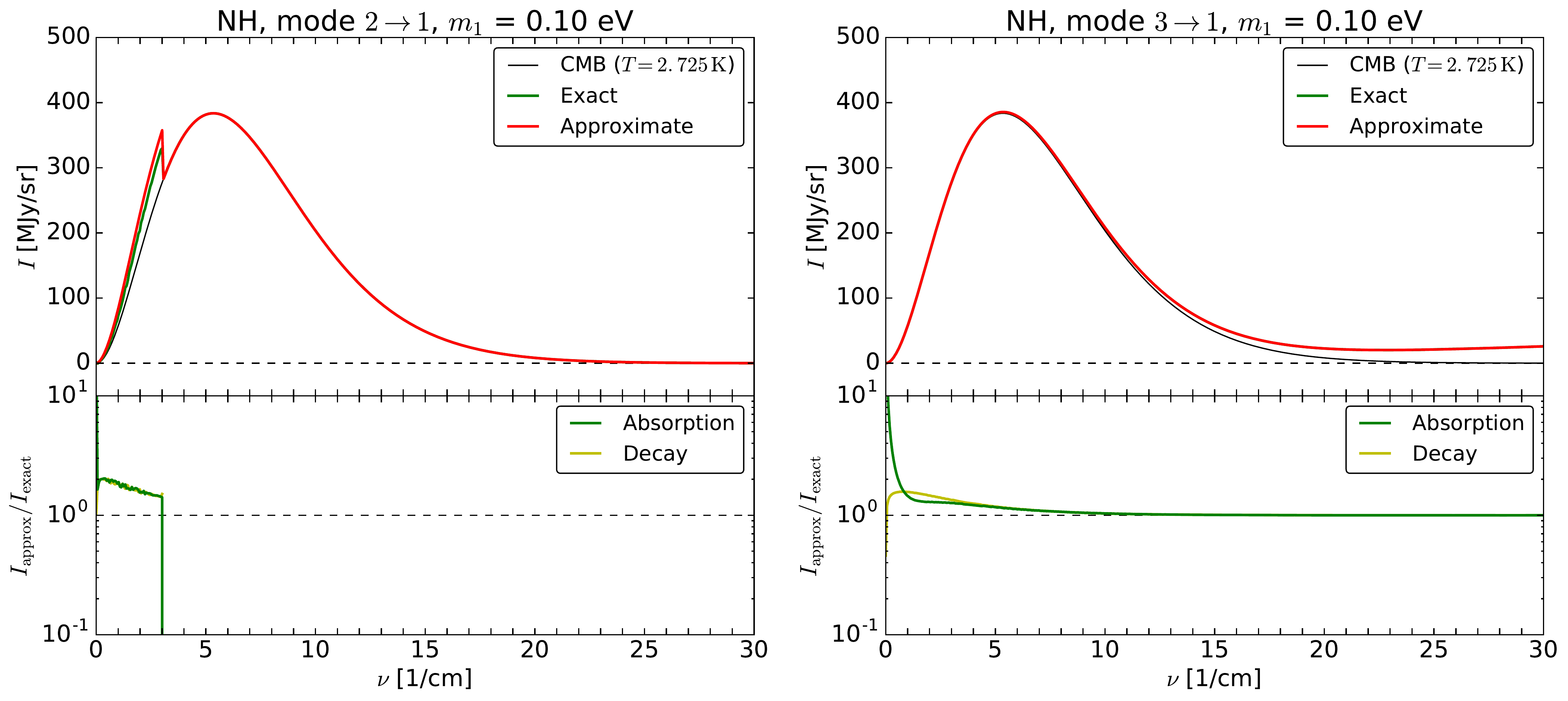}
    \caption{The same as Fig.~\ref{fig:ex_vs_approx_0.01} but for a heavier neutrino mass, $m_1 = 0.1$ eV.}
    \label{fig:ex_vs_approx_0.1}
\end{figure*}

Figures~\ref{fig:ex_vs_approx_0.01}~and~\ref{fig:ex_vs_approx_0.1}
compare the results of using the exact (green) and approximate (red)
calculations for different neutrino masses.
In each case, the left panels correspond to the decay/absorption of
$m_1$ and $m_2$, while the right panels show the same for $m_1$ and
$m_3$. 
Because $\Delta m_{13}^2$ is two orders of magnitude larger than $\Delta
m_{12}^2$, the spectral distortions for the $3 \to 1$ mode
extend up to much larger photon energies. 
From Fig.~\ref{fig:ex_vs_approx_0.01}, we see the effect of the neutrino
momentum distribution, which leads to a smoother cut-off in the
intensity when exact formulae are used, compared to the sharp cutoff in
approximate approach. Furthermore, at low frequencies, corresponding to high absorption redshift, the approximate absorption intensity is much larger than the exact absorption intensity. This suppression of the absorption intensity is a manifestation of the non-zero neutrino temperature, as described in the previous paragraph. Though these appear to be minor corrections, the high precision of the
CMB spectral measurements means that these should be taken into account
to obtain accurate limits on the neutrino lifetime.

Another effect which is observed in Figs.~\ref{fig:ex_vs_approx_0.01}
and \ref{fig:ex_vs_approx_0.1} is the impact of Pauli blocking.
In the exact formalism [Eqs.~\eqref{eq:P decay exact} and \eqref{eq:P
abs exact}], the term $1-1/[e^{W_{\pm}(s,t,x,y)}+1]$, leads to a
suppression of the decay and absorption rates when the final neutrino
state is already occupied. 
This Pauli-blocking effect lowers the overall intensity; we see from the
lower panels of Figs.~\ref{fig:ex_vs_approx_0.01} and
\ref{fig:ex_vs_approx_0.1} that the approximate intensity is always
larger than the exact one, by around 50\%.
The stimulated emission, on the other hand, enhances the decay
intensity, but the effect quickly decreases from $\sim$50\% at
2~cm$^{-1}$ (the lowest frequency of the FIRAS measurement) to $\lesssim
3$\% at $>6$~cm$^{-1}$.
Therefore, the absorption and Pauli-blocking combined should give lower
intensities than were found with the formulae used in
Ref.~\cite{Mirizzi2007}.


\section{Analysis of the COBE-FIRAS data of CMB spectrum and lower
 limits on decay lifetime}
\label{sec:analysis}

\subsection{Maximum likelihood analysis}
\label{sec:likelihood}
COBE-FIRAS has precisely measured the CMB spectrum, in
order to constrain cosmological parameters~\cite{Fixsen:1996nj}. 
The model for the CMB intensity discussed there included the effects of temperature deviations, Galactic
contamination, chemical potential $\mu$, and $y$-distortion. 
Since we also include the decay and absorption, we consider an intensity
$I$ of the form
\begin{eqnarray}
    \label{eq:intensitymodel}
    I &=& I_0 + \Delta T \frac{\partial I_\nu}{\partial T} + \mu
    \frac{\partial I_\nu}{\partial \mu} + G_0 I_{gal} + y I_y
    \nonumber\\&&{}+
    \Gamma_{ij} \left(I_{ij}^{\rm dec} +I_{ij}^{\rm abs} \right),
\end{eqnarray}
where the derivatives are to be evaluated at $T = T_0 = 2.725$~K and
$\mu = 0$, and $\Delta T \equiv T-T_0$.
The index $ij$ denotes the decay/absorption mode between the different
mass eigenstates of the neutrino: $ij \in \{12,13,23\}$. 
Furthermore $G_0$ is the amplitude of the Galactic contamination, $y$ is
the Kompaneets $y$ parameter, and $\Gamma_{ij}$ is the decay rate for
mode $ij$. 
$I_0$ is a regular black-body spectrum:
\begin{equation}
    I_0 = I_\nu (T,\mu) \big \rvert_{T=T_0, \mu=0} = \frac{2 h \nu^3}{e^{h \nu/T+\mu}-1} \bigg \rvert_{T=T_0, \mu=0},
\end{equation}
$I_{gal}$ is the residual Galactic contamination measured by FIRAS,
$I_y$ is given by \cite{1969Ap&SS...4..301Z, Fixsen:1996nj}
\begin{equation}
    I_y = T_0\left[\frac{h \nu}{T_0} \coth\left(\frac{1}{2}\frac{h \nu}{T_0}\right) -4 \right] \frac{\partial I_\nu}{\partial T} \bigg \rvert_{T=T_0,\mu=0},
\end{equation}
and $I_{ij}^{\rm dec}$, $I_{ij}^{\rm abs}$ are intensities corresponding
to decay and absorption, respectively, of mode $ij$ per unit
$\Gamma$.\footnote{Note that the intensities here are defined as
quantities per unit {\it frequency} range, instead of per unit {\it
energy} range as we defined in the previous section. We therefore have
to multiply the equations in Sec.~\ref{sec:formulation} by the Planck
constant $h = 2\pi$ to compute the intensities directly compared with
the FIRAS data.}
Note that in reality, all three modes occur simultaneously, but we find
that including them all at once in the analysis would yield
unnecessarily weak constraints on $\Gamma_{ij}$. 
This is because the masses of $\nu_1$ and $\nu_2$ are relatively close,
resulting in degeneracy in the spectra of the modes 13 and 23. 
We solve this issue by focusing on one mode at a time, forcing the other
two decay rates to be zero.

The parameters we are interested in are $\Gamma_{12}$, $\Gamma_{13}$ and
$\Gamma_{23}$. 
We will constrain these by fitting the model in
Eq.~\eqref{eq:intensitymodel} to the FIRAS data, and minimizing $\chi^2$
as a function of the parameters $\Delta T$, $\mu$, $G_0$, $y$, and
$\Gamma_{ij}$. 
The $\chi^2$ of this model is given by 
\begin{equation}
    \chi^2 = \sum_{i,j=1}^{43} (I^{\rm data}_i-I^{\rm model}_i)
     (C^{-1})_{ij} (I^{\rm data}_j-I^{\rm model}_j),
\end{equation}
where $I^{\rm model}$ is given by Eq.~\eqref{eq:intensitymodel}, $I^{\rm
data}$ is the FIRAS measurement, and $C$ is the covariance matrix taken
from Ref.~\cite{Fixsen:1996nj}. 
The sum runs over the 43 frequency bins of FIRAS.



Our model for the intensity is defined by the parameters $\theta_a \in \{\Delta T, \mu, G_0, y,
\Gamma_{ij}\}$, whose best fit values $\hat{\theta}_a$ are determined by solving the system of 5 simultaneous equations:
\begin{equation}
\label{eq:bestfit}
   \left.\frac{\partial \chi^2}{\partial
    \theta_a}\right|_{{\hat{\bm\theta}}} = 0.
\end{equation}
In order to estimate errors for each parameter $\theta_a$, taking
degeneracy among the parameters into account, we calculate the Observed Fisher
Information matrix:
\begin{equation}
\label{eq:fisher}
    F_{ab} = \frac{1}{2} \frac{\partial^2 \chi^2}{\partial \theta_a
     \partial \theta_b}.
\end{equation}
The parameters $\theta_a,\, \theta_b \in \{\Delta T, \mu, G_0, y,
\Gamma_{ij}\}$ so $\dim(F) = 5$, since we are only looking at one decay
mode at a time. The derivatives in Eq.~\eqref{eq:fisher} are to be
evaluated at the best-fit point $\hat{\theta}_a$.
However, we note that assuming the linearized intensity in
Eq.~\eqref{eq:intensitymodel}, the Fisher Information is independent of
the parameters $\theta_a$ and $\theta_b$.

The covariance between parameters $\theta_a$ and $\theta_b$ is the
inverse of this matrix:
\begin{equation}
\label{eq:correlation coefficient}
    \mbox{Cov}(\theta_a,\theta_b) = \left(F^{-1}\right)_{ab}\,.
\end{equation}
The $1\sigma$ uncertainty of a specific parameter $\theta_a$ is equivalent to the
diagonal components of the covariance matrix as follows:
\begin{equation}
\label{eq:sigma}
    \sigma_{a} = \sqrt{\left(F^{-1}\right)_{aa}}\,.
\end{equation}
The upper limit at 95\% confidence level (C.L.) on the parameter
$\theta_a$ (corresponding to $\Delta \chi^2 \approx 2.71$) is then
estimated as
\begin{equation}
    \theta_a^{95\%} \approx \hat{\theta}_a + \sqrt{2.71}\, \sigma_a\,.
\end{equation}
In some cases, we find that the best-fit value, $\hat{\Gamma}$, is
negative, which is clearly unphysical.
In this case, we assume that Eq.~\eqref{eq:fisher} remains a good
approximation to the $\chi^2$.
The physical best-fit is then at $\Gamma = 0$ and we calculate the upper
limit as:
\begin{equation}
    \Gamma^{95\%} \approx \hat{\Gamma} + \sqrt{\hat{\Gamma}^2 + 2.71 \,\sigma_\Gamma^2}\,.
\end{equation}

We have checked our analysis procedure by fixing $\Gamma = 0$ and determining limits on the $\mu$ and $y$ paramater separately. Our results are consistent with those reported in Ref.~\cite{Fixsen:1996nj} ($| \mu | < 9 \times 10 ^ { - 5 }$ and $|y| < 1.5 \times 10^{-5}$ at 95\% CL).

\subsection{Constraints on neutrino decay lifetime and transition
  magnetic moments}
\label{sec:MagneticMoment}

\begin{figure*}[t]
    \centering
    \includegraphics[width=0.49\textwidth]{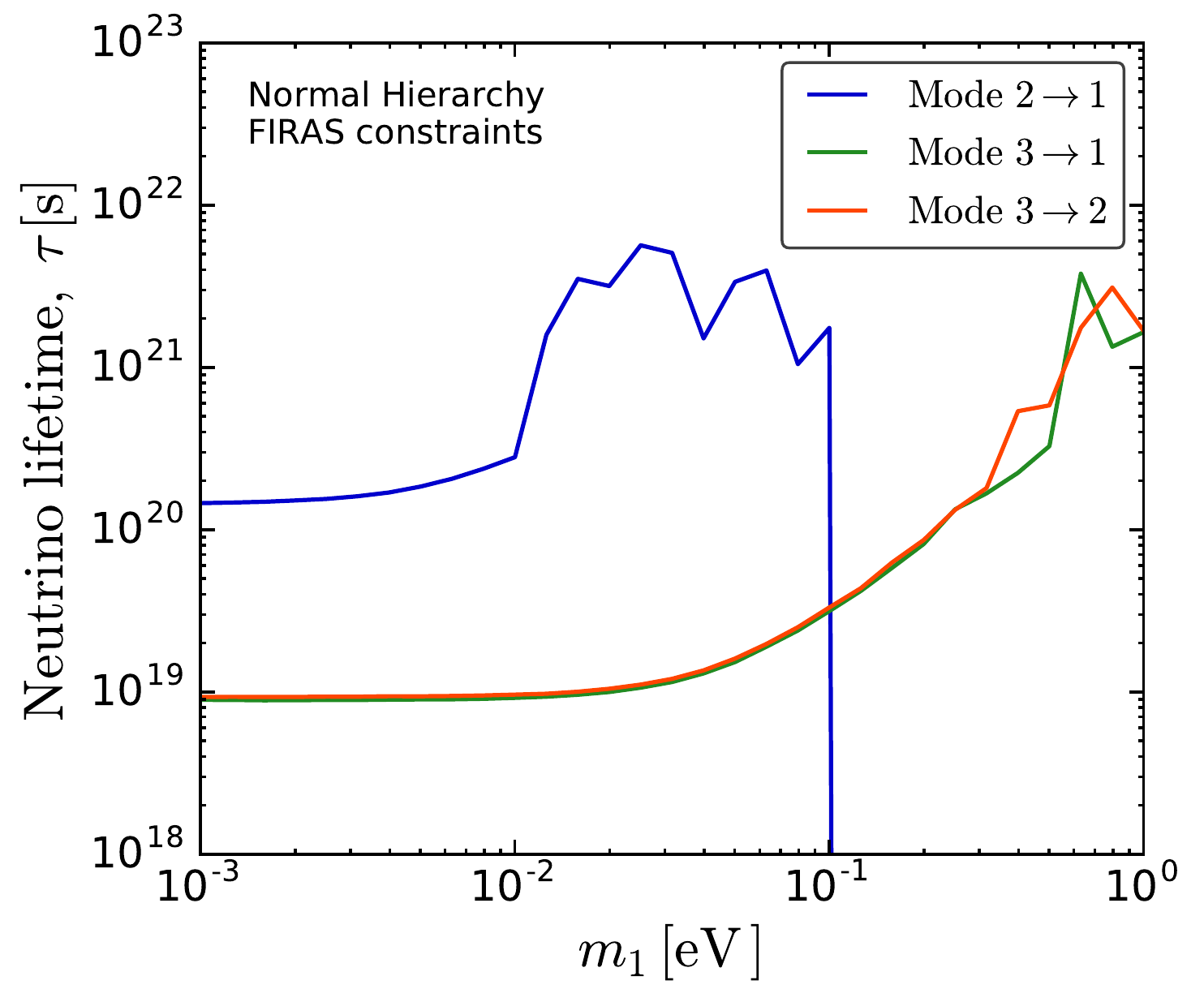}
        \includegraphics[width=0.49\textwidth]{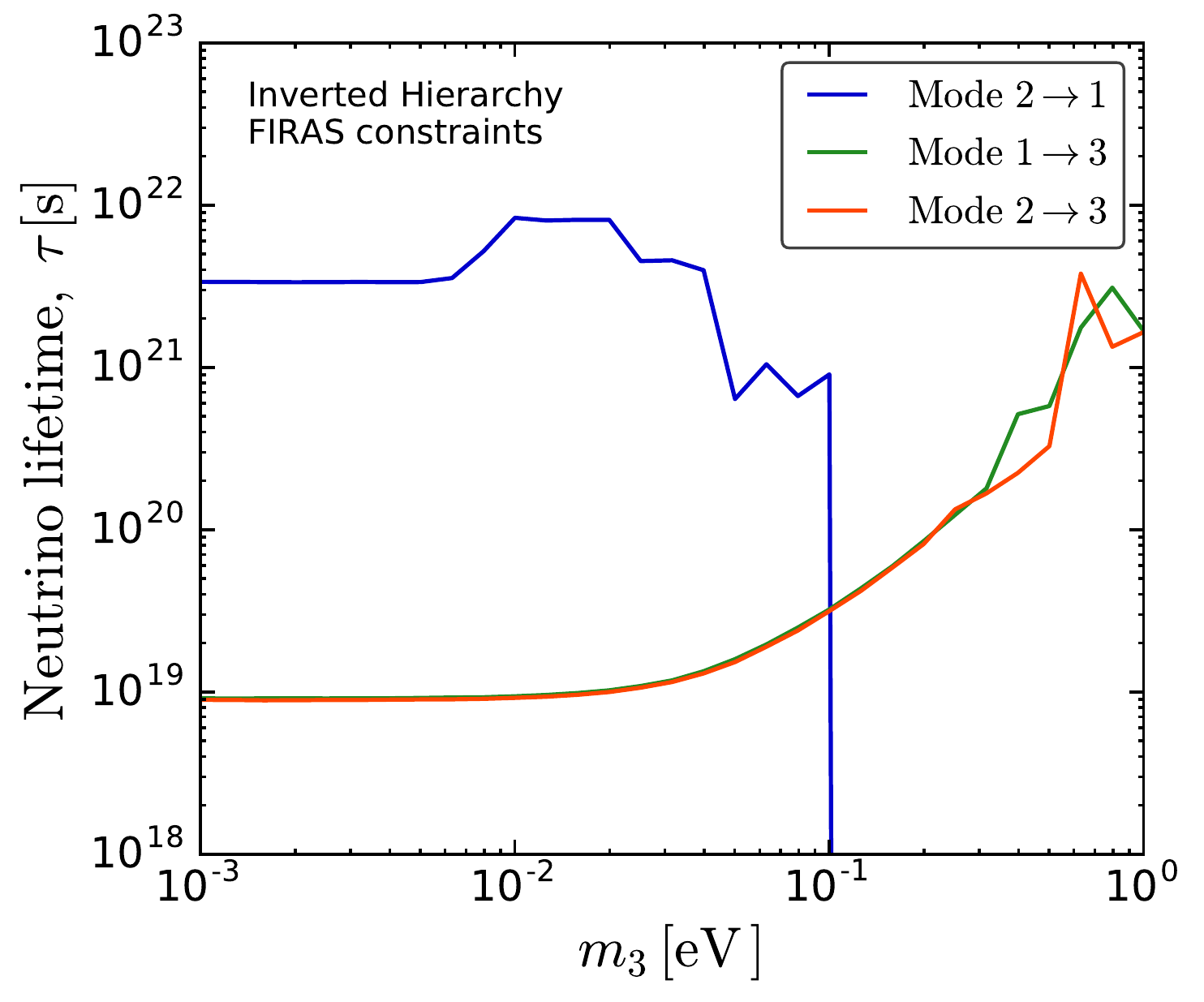}
    \caption{\textbf{95\% C.L. lower limits on radiative decay lifetime of neutrinos as a function of the lightest neutrino mass}. Values of the neutrino lifetime below the 
 solid curves are excluded by our analysis at the 95\% C.L. \textbf{Left panel:} Results
 for NH, where $m_1$ is the lowest mass.  \textbf{Right panel:} Results
 for IH where $m_3$ is the lowest mass.}
    \label{fig:tau}
\end{figure*}

Using the approach given in the previous subsection, and the FIRAS
data~\cite{Fixsen:1996nj}, we numerically compute values for the 95\% C.L. lower limit on the neutrino lifetime $\tau = 1/\Gamma$ as a function of the lowest neutrino mass.
These constraints are presented in Fig.~\ref{fig:tau} for NH (left
panel) and IH (right panel).

Constraints on the 13 and 23 modes are weaker than for the 12 mode by
around an order of magnitude. This is as expected comparing, e.g., the
left and right panels of Fig.~\ref{fig:ex_vs_approx_0.01},
where the distortion due to the 12 mode is clearly larger for a fixed
value of $\tau$.
As outlined in Sec.~\ref{sec:theory results}, this is because the
larger mass squared difference in the 13 case means that the majority
of the distortion appears at frequencies above the FIRAS range.
For the 12 mode, the strongest constraints appear in the mass range
0.01--0.12~eV, below which the sharp spectral feature from neutrino
decay lies above the FIRAS frequency range.
Above $\sim$0.12~eV, the CMB distortions from neutrino decay and absorption
occur at too low frequencies to be detected by FIRAS.
As pointed out in Ref.~\cite{Mirizzi2007}, the jagged shape of the
limits is due to the fact that the $\chi^2$ changes abruptly when the
end-point of the neutrino decay spectrum crosses into a new frequency
bin.

\begin{figure*}[t]
    \centering
    \includegraphics[width=0.49\textwidth]{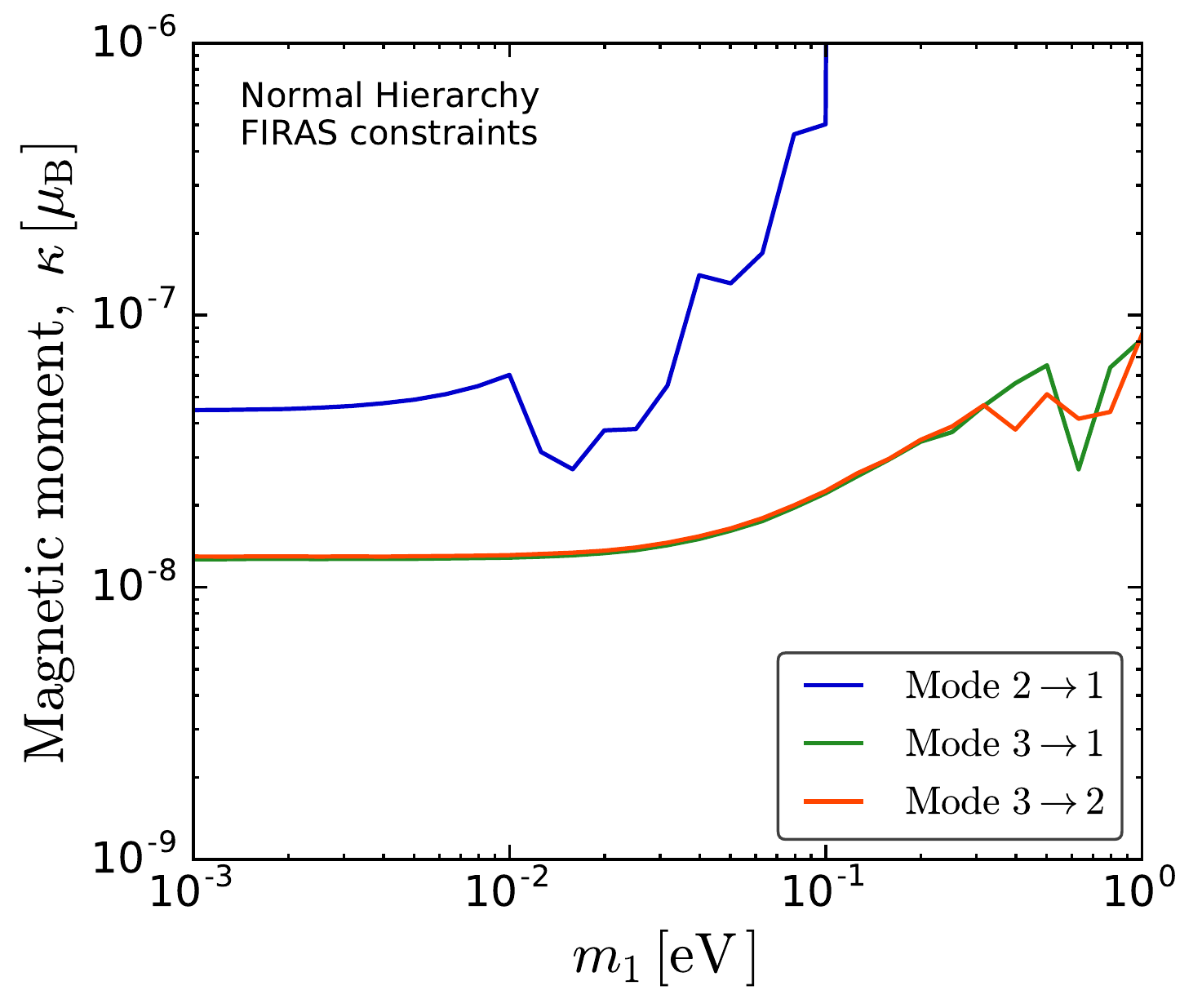}
     \includegraphics[width=0.49\textwidth]{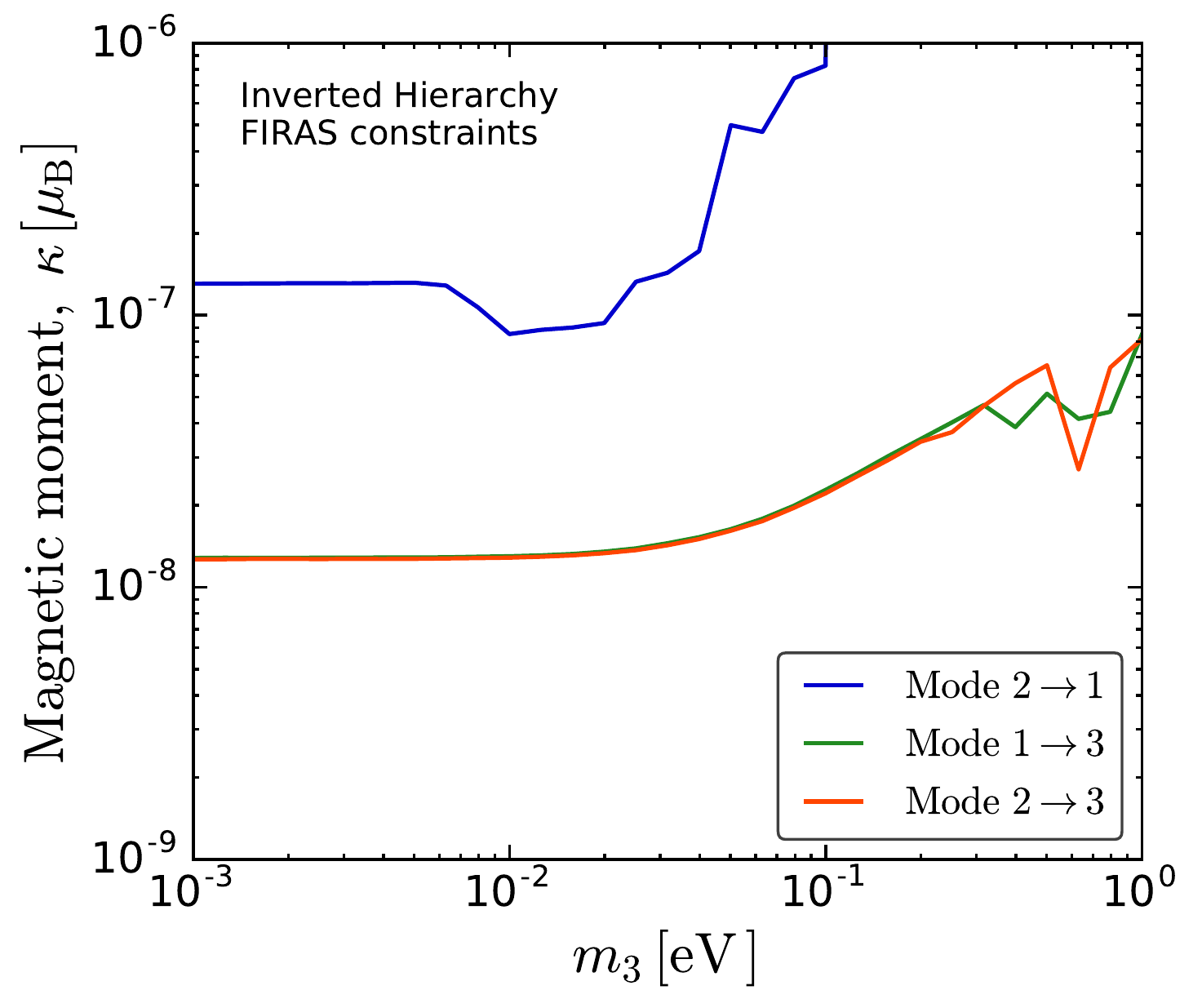}
    \caption{\textbf{95\% C.L.~upper limits on the magnetic moment of neutrinos, as a function of the lightest neutrino mass.} Values of the effective transition magnetic moment $\kappa_{ij}$ [defined in Eq.~\eqref{eq:kappa}] above the solid lines are excluded by our analysis. Results shown are for NH (\textbf{left panel}) and IH (\textbf{right panel}).}
    \label{fig:magmom}
\end{figure*}

We now translate our constraints on the radiative neutrino decay rate
$\Gamma$ into constraints on the effective neutrino magnetic moment.
For neutrinos with transition magnetic and electric moments, $\mu_{ij}$
and $\epsilon_{ij}$, respectively, we can define the effective magnetic
moment $\kappa_{ij}^2 \equiv |\mu_{ij}|^2 + |\epsilon_{ij}|^2$.
For a transition  $\nu_j \rightarrow \nu_i + \gamma$, the decay rate
induced by this magnetic moment is given by~\cite{Raffelt:1996wa}:
\begin{equation}
\label{eq:kappa}
 \Gamma_{ij} = \frac{\kappa_{ij}^2}{8\pi}
 \left(\frac{\Delta m_{ij}^2}{m_j}\right)^3 .
\end{equation}
The resulting constraints on $\kappa_{ij}$ are shown in
Fig.~\ref{fig:magmom}.
In the case of NH (left panel), our constraints extend down to
$\alt 10^{-8} \;\mu_B$ for the 13 and 23 modes and $\alt 4\times
10^{-8} \;\mu_B$ for the 12 mode.
In the case of IH (right panel), the constraints on the 12 mode are
weaker by roughly a factor of 2.
This is because $m_1$ and $m_2$ are larger than the case in NH, leading
to a smaller decay rate for a given magnetic moment
[Eq.~\eqref{eq:kappa}].
This dependence of the decay rate on the neutrino mass also explains why
the 23 and 13 modes give stronger constraints on $\kappa_{ij}$ than the
12 mode (the opposite was seen in Fig.~\ref{fig:tau}).

\subsection{Comparison to earlier work and degeneracy among parameters}

We now compare our results to those previously obtained by
Ref.~\cite{Mirizzi2007}.
The analysis of Ref.~\cite{Mirizzi2007} did not take into account
stimulated emission, Pauli-blocking or absorption, and assumed the
neutrinos to be at rest at the moment of decay.
In addition, only $\Gamma$ was varied in the $\chi^2$ analysis.
The difference between these results should tell us the impact of the
exact calculation on the CMB spectrum as well as the importance of
including additional nuisance parameters in the analysis.

When we compare our exact results to the findings in
Ref.~\cite{Mirizzi2007}, we find that our results are in broad agreement
with the results of Fig.~2 presented there.
For NH, we obtain a stronger limit for the 12 mode in the range of $
10^{-2} \;\mathrm{eV} \lesssim m_{\nu} \lesssim 10^{-1} \;\mathrm{eV}$
by about one order of magnitude.
For IH, our bounds on the 13 and 23 modes are slightly weaker (by a
factor of around 2) than the bounds found in Ref.~\cite{Mirizzi2007} in
the region $10^{-3} \;\mathrm{eV} \lesssim m_{\nu} \lesssim 10^{-1}
\;\mathrm{eV}$. We again obtain a stronger bound on the 12 mode in the
region $ 10^{-2} \;\mathrm{eV} \lesssim m_{\nu} \lesssim 10^{-1}
\;\mathrm{eV}$.
We emphasize that we expect our constraints to be more accurate, as we include more accurate calculations of the spectral distortions and more parameters in the analysis.

To further investigate how our results differ from the previous
constraints, we have repeated our analysis, following (where possible)
the analysis procedure of Ref.~\cite{Mirizzi2007}.
In order to do this, we use the following model for the intensity:
\begin{equation}
    I = I_0 + \Gamma_{ij} I_{ij}^{\mathrm{dec}},
\end{equation}
instead of the full model given in Eq.~\eqref{eq:intensitymodel}, fixing
all the other parameters to be zero.
Note that we do not include the contribution of absorption or stimulated
emission, and calculate the decay intensity using the approximate
approach presented in Sec.~\ref{sec:approximate formulae}.
The uncertainty on the decay rate is then given by $\sigma_\Gamma^2 =
2(\partial^2 \chi^2/\partial \Gamma^2)^{-1}$ (not taking into account
the full Fisher matrix).
We also adopt the mass differences and cosmological parameters stated in
Ref.~\cite{Mirizzi2007}.

\begin{figure*}[t]
    \centering
    \includegraphics[width=0.49\textwidth]{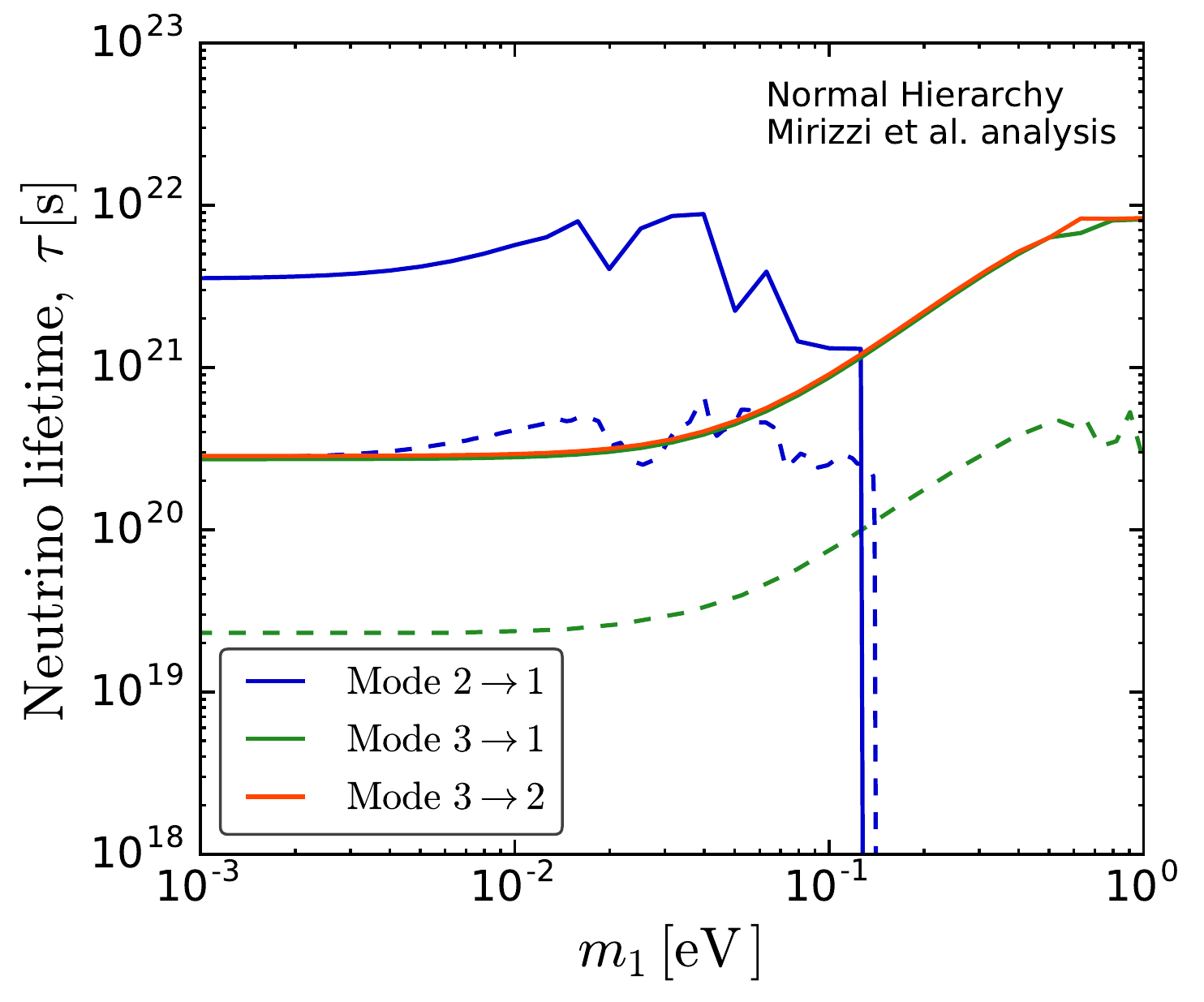}
        \includegraphics[width=0.49\textwidth]{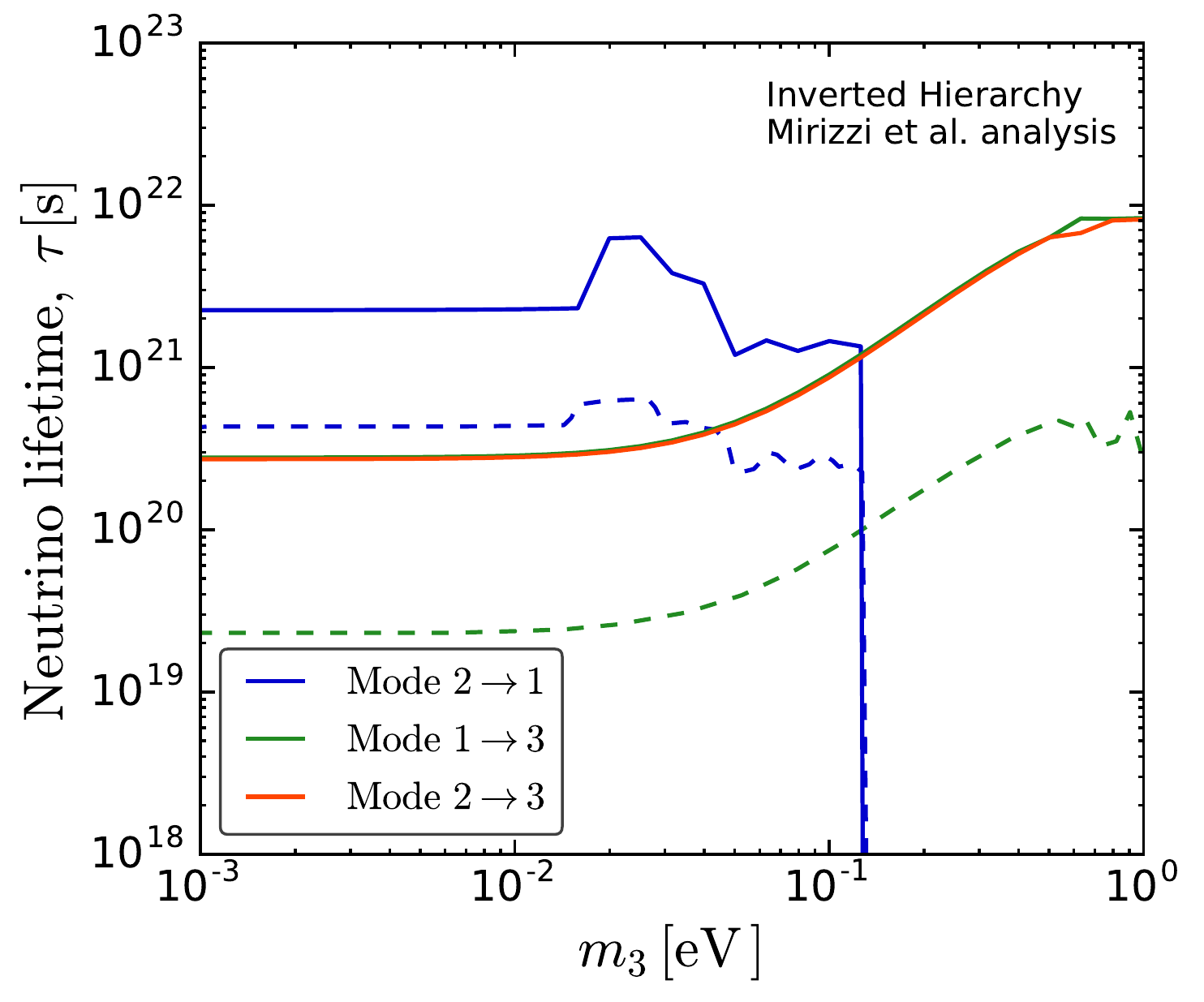}
    \caption{\textbf{95\% C.L.~lower limits on the radiative decay lifetime of neutrinos, following the analysis method of Ref.~\cite{Mirizzi2007}.} These constraints were derived using approximate expressions for neutrino decay only and fixing all nuisance parameters except for $\Gamma$. Dashed lines are the limits reported by Ref.~\cite{Mirizzi2007}. Results shown are for NH (\textbf{left panel}) and IH (\textbf{right panel}).}   \label{fig:tau_constraints_NH_approx}
\end{figure*}

The resulting lower limits on $\tau$ are shown in
Fig.~\ref{fig:tau_constraints_NH_approx}, where dashed lines are the
bounds reported in Ref.~\cite{Mirizzi2007} and solid lines show our
bounds using the same analysis approach.
The shape of the bounds is in close agreement.
However, we notice that our bounds are around one order of magnitude
stronger, although we have attempted to reproduce the analysis of
Ref.~\cite{Mirizzi2007} as closely as possible.
Unfortunately, we have not been able to find the source of this
discrepancy.\footnote{Reference~\cite{Mirizzi2007} defines a reduced
chi-squared test statistic, but do not specify how they calculate upper
limits from this, so it is difficult to reproduce their bounds exactly.}

\begin{figure*}
 \begin{center}
  \includegraphics[width=8.5cm]{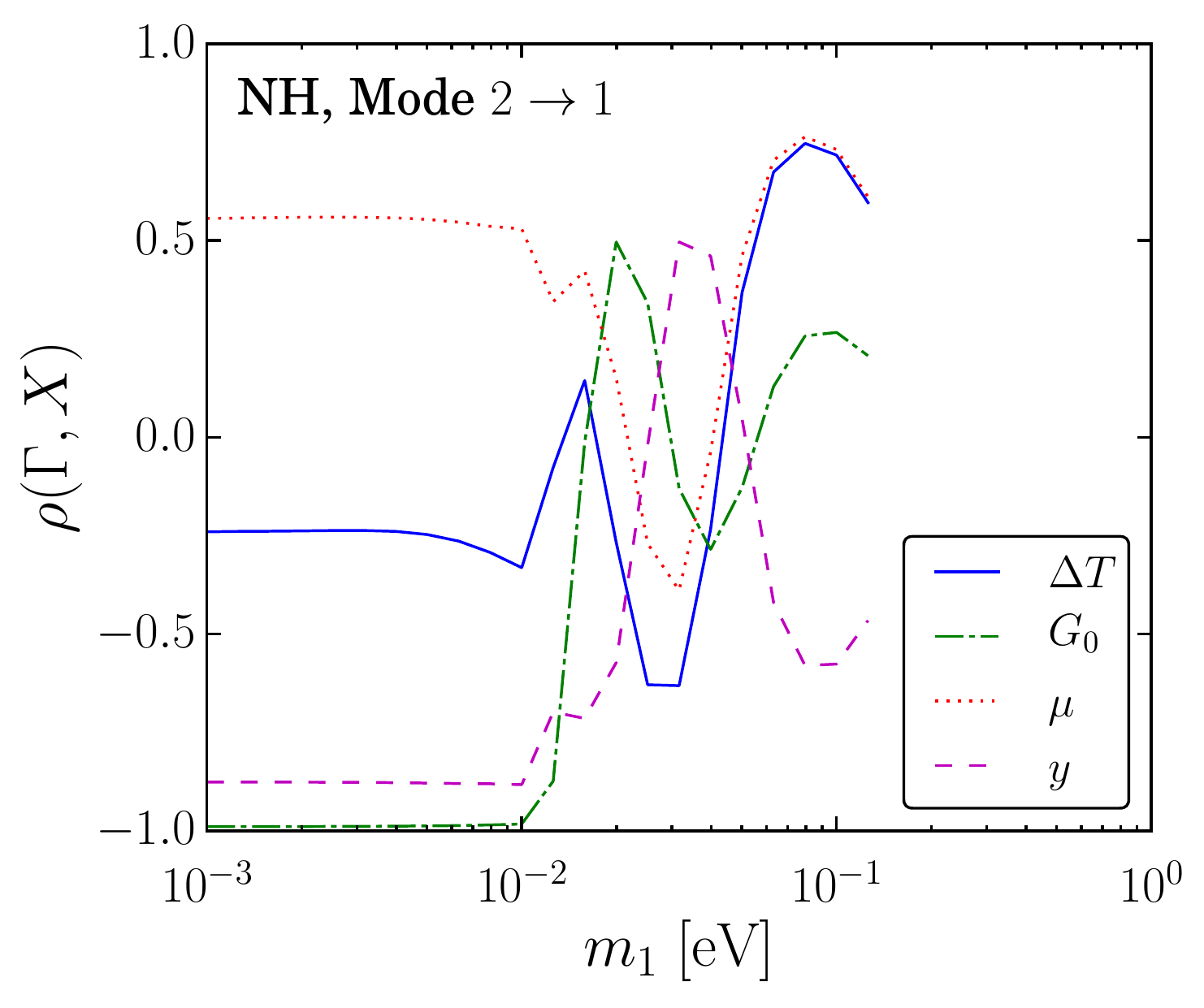}
  \includegraphics[width=8.5cm]{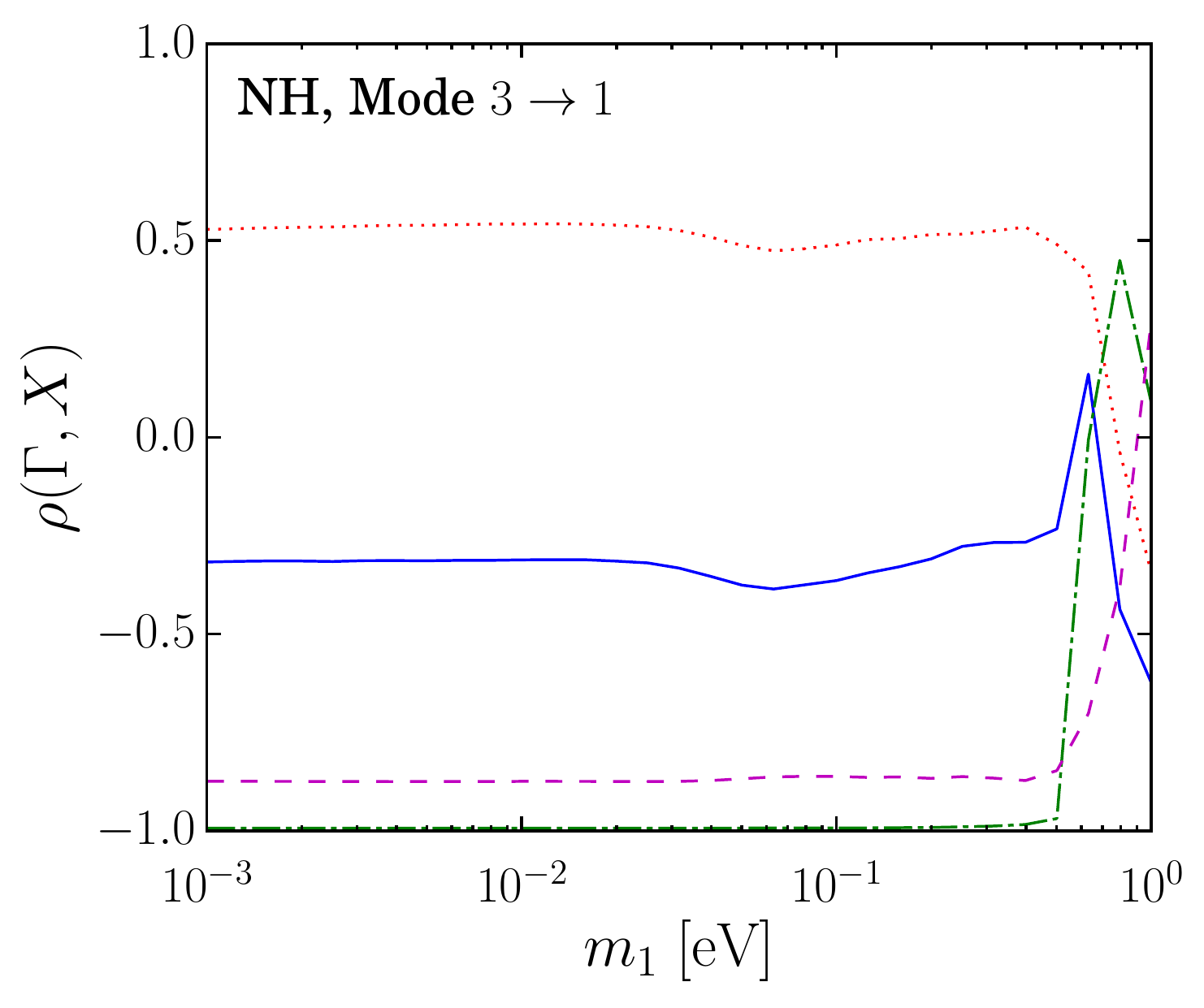}
  \includegraphics[width=8.5cm]{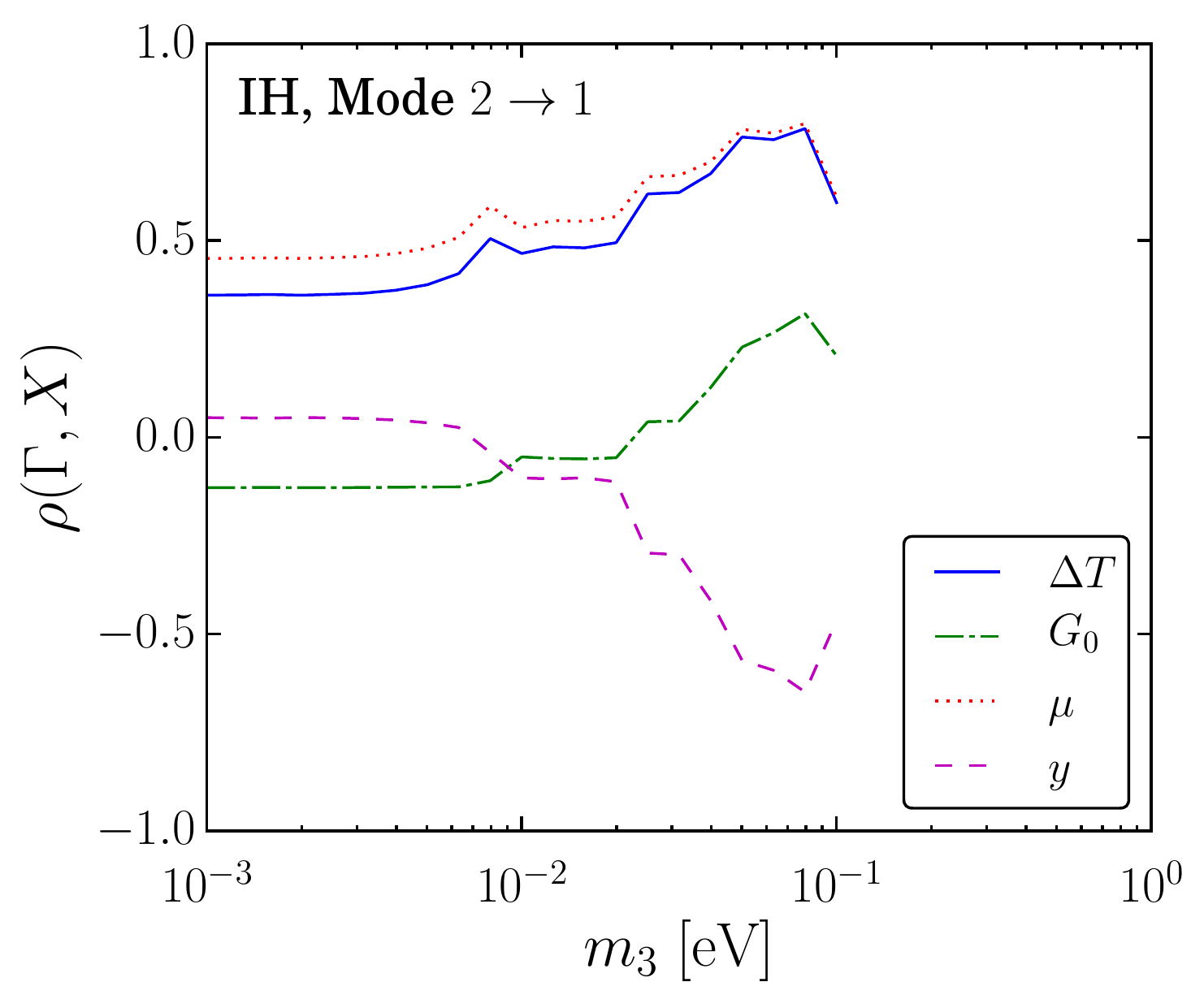}
  \includegraphics[width=8.5cm]{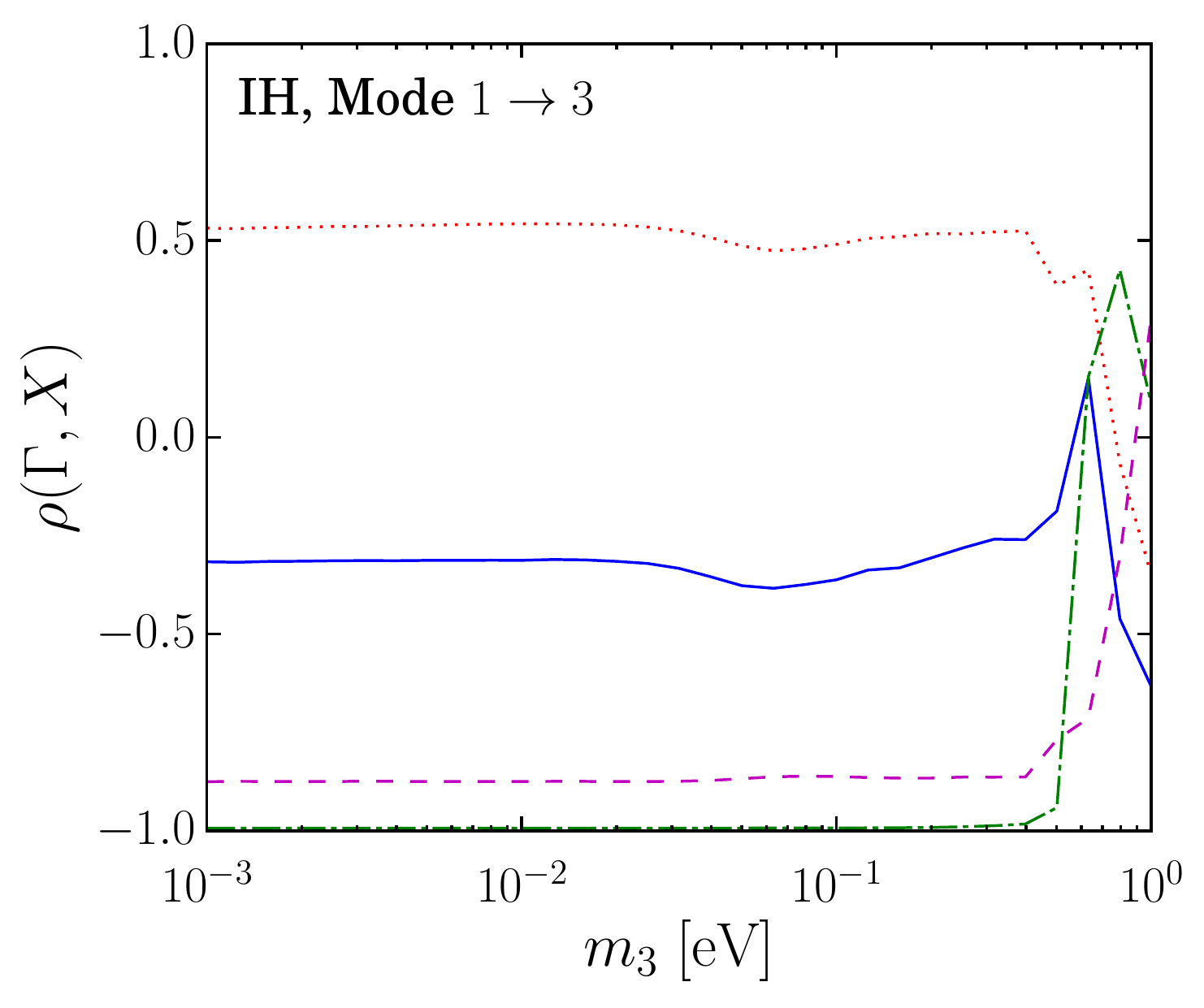}
  \caption{\textbf{Correlation coefficients $\rho(\Gamma, X)$
   between the decay
  rate $\Gamma$ and other parameters $X \in \{\Delta T, G_0, \mu, y\}$.}
  The correlation coefficients [Eq.~(\ref{eq:correlation coefficient})]
  are shown as a function of the lightest neutrino mass, for NH
  (\textbf{top row}) and IH (\textbf{bottom row})
  and for the modes 12 (\textbf{left column}) and 13 (\textbf{right column}).}
  \label{fig:corr_coeff}
 \end{center}
\end{figure*}

Lastly, when comparing our approximate bounds in
Fig.~\ref{fig:tau_constraints_NH_approx} with the full analysis from
Fig.~\ref{fig:tau}, we notice that the full bounds are typically weaker
by a factor of $\sim$30.
This implies that we cannot ignore the correlation among different
parameters (which are included in the full analysis).
To see the effect more quantitatively, we introduce the correlation
coefficients:
\begin{equation}
 \rho(\theta_a,\theta_b) \equiv
  \frac{\mbox{Cov}(\theta_a,\theta_b)}{\sigma_a\sigma_b},
  \label{eq:correlation coefficient}
\end{equation}
and show them between $\Gamma_{ij}$ and the other parameters in
Fig.~\ref{fig:corr_coeff} for the modes 12 and 13 and for both the NH
and IH (those for the mode 23 is almost identical to those for 13).
In fact, we find very strong anti-correlation between $\Gamma_{13}$ (or
$\Gamma_{23}$) and the Galactic component $G_0$ as well as the
$y$-distortion for nearly all the masses investigated here.
This is because the modification of the CMB spectrum increases as a
function of the frequency without any feature, and is thus
indistinguishable from the Galactic residual component found in
Ref.~\cite{Fixsen:1996nj} up to the FIRAS errors.
On the other hand, if a sharp spectral feature appears in the FIRAS
frequency range (as is the case for $m_1 \in[10^{-2},
10^{-1}]\;\mathrm{eV}$ in the 12 mode), it breaks the degeneracy and
hence the anti-correlation disappears (upper left panel of
Fig.~\ref{fig:corr_coeff}).
Indeed, we see that comparing the full (Fig.~\ref{fig:tau}) and
approximate analysis (Fig.~\ref{fig:tau_constraints_NH_approx}), the
bounds in this mass range are largely unchanged between the two.

These results show that in the full analysis, including $\theta_a \in \{\Delta T, \mu, G_0, y,
\Gamma_{ij}\}$, degeneracies between the parameters can substantially weaken constraints on the neutrino decay rates $\Gamma_{ij}$. While accounting for these degeneracies represents the most conservative approach, we could alternatively have chosen to fix $\mu = 0$ and $y = 0$ in the analysis. In the standard $\Lambda$CDM cosmology, we expect $y \in [10^{-7}, 10^{-6}]$, caused by heating during reionization and other heating mechanisms \cite{Hu:1993tc,Refregier:2000xz,Oh:2003sa} and $\mu \sim \mathcal{O}(10^{-8})$, from the damping of primordial fluctatuations \cite{Hu:1992dc}. These values lie below the FIRAS sensitivity and so, if we assume no other sources of $\mu$- and $y$-distortions, we could keep these parameters fixed (effectively to zero) in the analysis. The solid lines in Fig.~\ref{fig:tau_constraints_NH_approx} give an estimate of the limits on $\Gamma_{ij}$ in this case. As we discuss in the next section, future experiments will be more sensitive to $\mu$ and $y$, in which case their inclusion in the analysis is unavoidable.

       

\section{Sensitivity of future CMB experiments}
\label{sec:future}

Highly sensitive future CMB measurements will be able to measure
spectral distortions to a high degree of precision.
Of particular interest are  $\mu$- and $y$-distortions, briefly discussed in the previous section. These provide
information about energy-release at certain redshifts and therefore
allow us to constrain the thermal history of the
Universe~\cite{Chluba2}.
A measurement of $y$-distortions may provide information about structure
formation and the epoch of reionization at $z < 10$--20, as well as
allowing us to probe the primordial power spectrum on small
scales~\cite{1970Ap&SS...7....3S}.
The decay and annihilation of particles in the pre-recombination epoch
($5\times 10^4 < z < 2\times 10^6$) may give rise to
$\mu$-distortions~\cite{Hu1993}, providing sensitivity to particle
lifetimes in the range $\tau \simeq 10^{8}$--$10^{11}$ s.
As we have explored so far in this work, particles with longer lifetimes
may also distort the CMB spectrum and provide a detectable signal in
future CMB experiments.

Here we focus on the PIXIE mission which is expected to cover a
frequency range of 30~GHz (1~cm$^{-1}$) to 6~THz (200~cm$^{-1}$), using
400 channels.
For this analysis, however, we will only look at a range 30--750~GHz
(1--25~cm$^{-1}$, divided into 48 frequency bins), as most of the CMB
spectrum lies within this range.
Furthermore, at frequencies higher than 3~THz, the spectrum is dominated
by dust emitting foregrounds that do not affect the final
analysis~\cite{Abitbol2017}.

Following closely the analysis of Sec.~\ref{sec:analysis}, we obtain
projected lower limits on the neutrino lifetime $\tau$ from the PIXIE
experiment.
We assume that the error on the intensity in each frequency bin is
5~Jy/sr~\cite{KogutPixie}, and that there are no correlations between
the different frequency bins.
We include the parameters $ \theta_a \in \{\Delta T, \mu, G_0, y,
\Gamma_{ij}\}$ in the modeled intensity, with the projected Galactic
contamination taken from Ref.~\cite{Abitbol2017}.
We also consider the ideal case in which $G_0$ is fixed to zero;
i.e., the Galactic contamination, presumably calibrated with other
wavebands, is well constrained and perfectly subtracted.

Unlike in Sec.~\ref{sec:analysis}, the intensity spectrum has not yet been measured by PIXIE. We therefore assume that the best fit decay rate will be $\hat{\Gamma} = 0$. Using the Fisher-matrix approach of Sec.~\ref{sec:analysis}, we then estimate
the 95\% C.L. projected limit\footnote{We might also call this the projected \textit{sensitivity} of PIXIE.} on $\Gamma$ as $\Gamma^{95\%}
\approx 1.64 \sigma_{\Gamma}$, where $\sigma_\Gamma$ is defined in
Eq.~\eqref{eq:sigma}. As noted in Sec.~\ref{sec:likelihood}, in our linearised intensity model the numerical value of the Fisher matrix does not depend on the model parameters $\theta_a$. This means that the projection we obtain for $\Gamma^{95\%}$ does not depend on the assumed values of the nuisance parameters (although it would depend on the assumed best fit of the decay rate $\hat{\Gamma}$).

\begin{figure*}
\centering
  \includegraphics[width=0.49\textwidth]{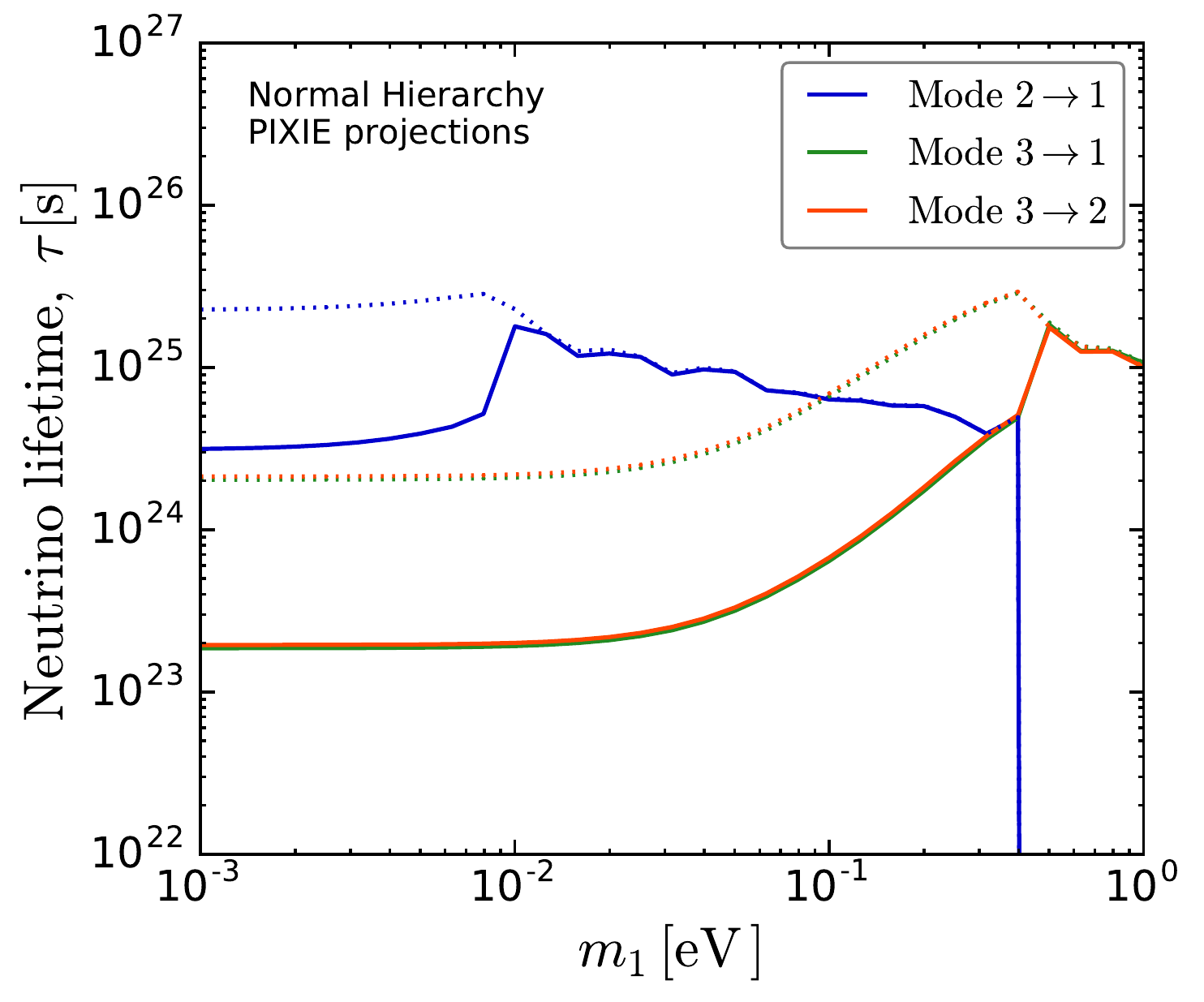}
    \includegraphics[width=0.49\textwidth]{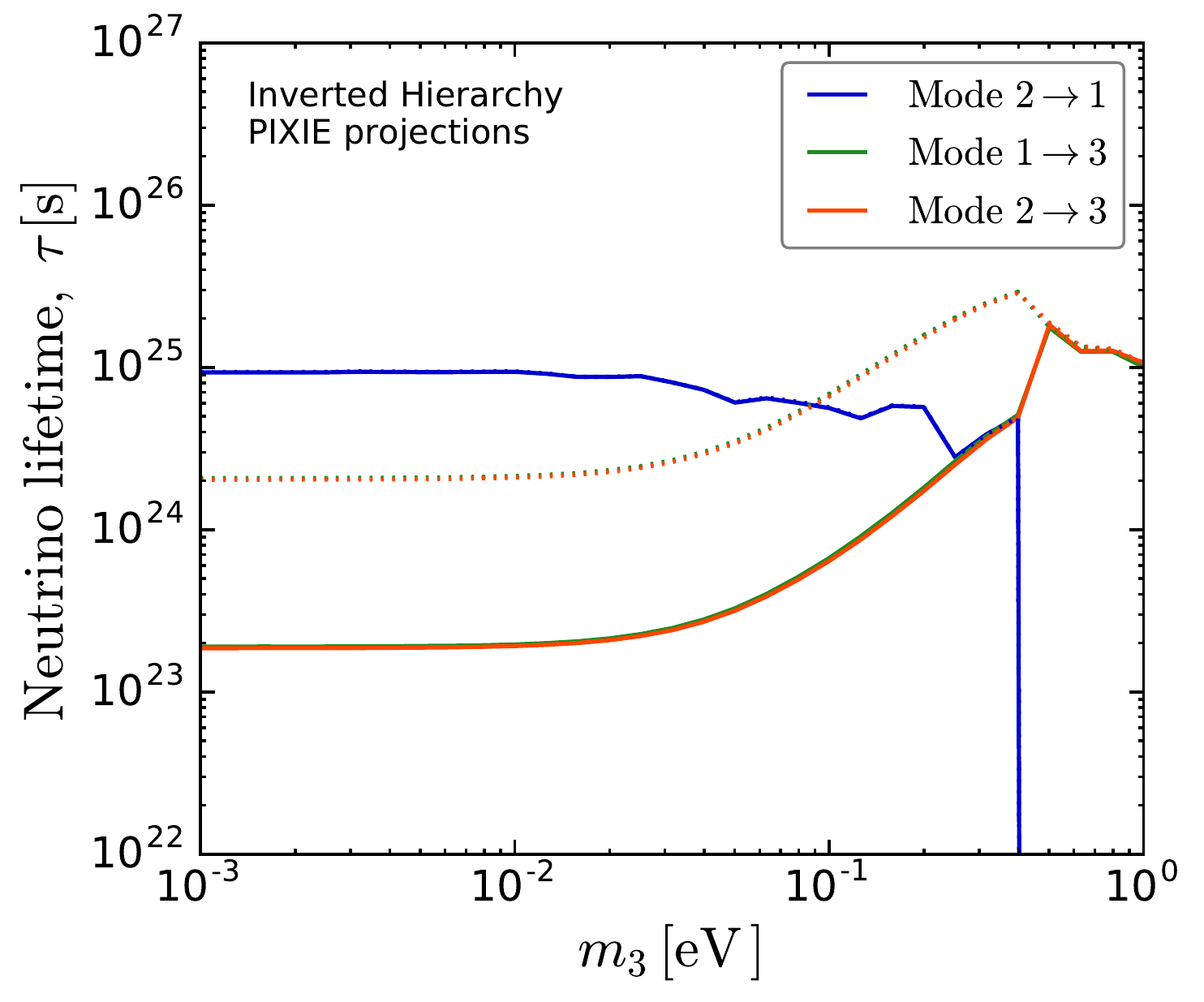}
\caption{\textbf{Projected 95\% C.L.~lower limits on radiative decay lifetime of neutrinos as a function of the lightest neutrino mass for a PIXIE-like experiment}. Dotted lines correspond to the lower limits assuming that the residual Galactic contamination $G_0$ is fixed to zero. \textbf{Left panel:} Results
 for NH, where $m_1$ is the lowest mass.  \textbf{Right panel:} Results
 for IH where $m_3$ is the lowest mass.}
\label{fig:GroupD tau}
\end{figure*}

The PIXIE projected limits are shown in Fig.~\ref{fig:GroupD tau}.
Solid lines show the projections including parameters $ \theta_a \in
\{\Delta T, \mu, G_0, y, \Gamma_{ij}\}$, while dotted lines show the
projection when $G_0$ is fixed to zero.
The qualitative behavior of the bounds matches those from FIRAS,
although for the 12 mode the bounds extend to higher values of $m_1$ as
PIXIE will probe down to lower frequencies than FIRAS.
The projected limits lie in the range $\tau \gtrsim 10^{23}$--$10^{25}$
s, representing a factor of $10^4$ improvement over the FIRAS limits. This improvement arises both from a reduction of the uncertainties on the CMB intensity and from an increase in the number of frequency channels from FIRAS to PIXIE. 
Fixing the Galactic component to zero improves the constraints by
roughly another order of magnitude (unless the spectral feature lies
within the PIXIE frequency range, as is the case for masses above
$10^{-2}$ eV in the 12 mode).

\begin{figure*}
\centering
  \includegraphics[width=0.49\textwidth]{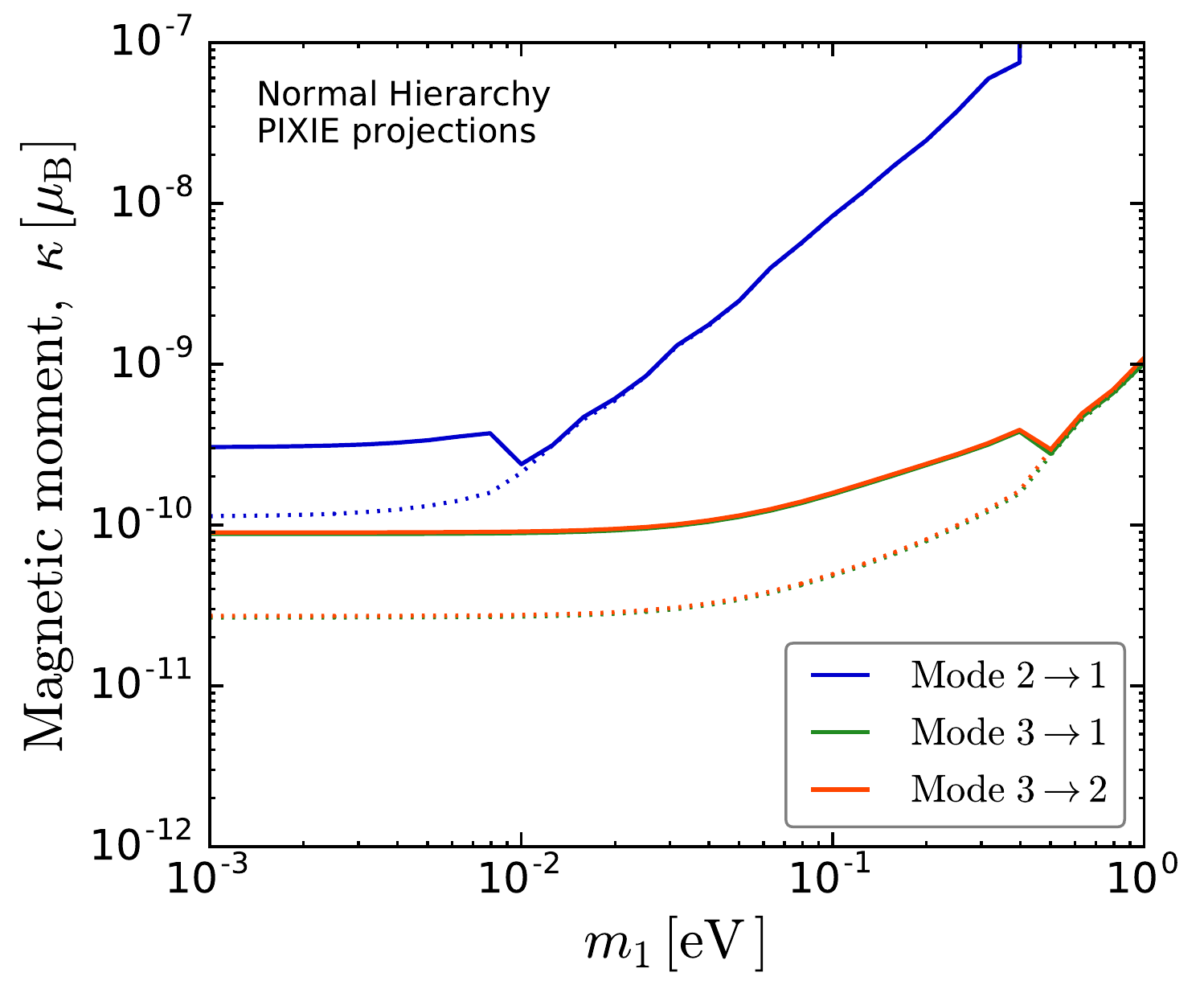}
  \includegraphics[width=0.49\textwidth]{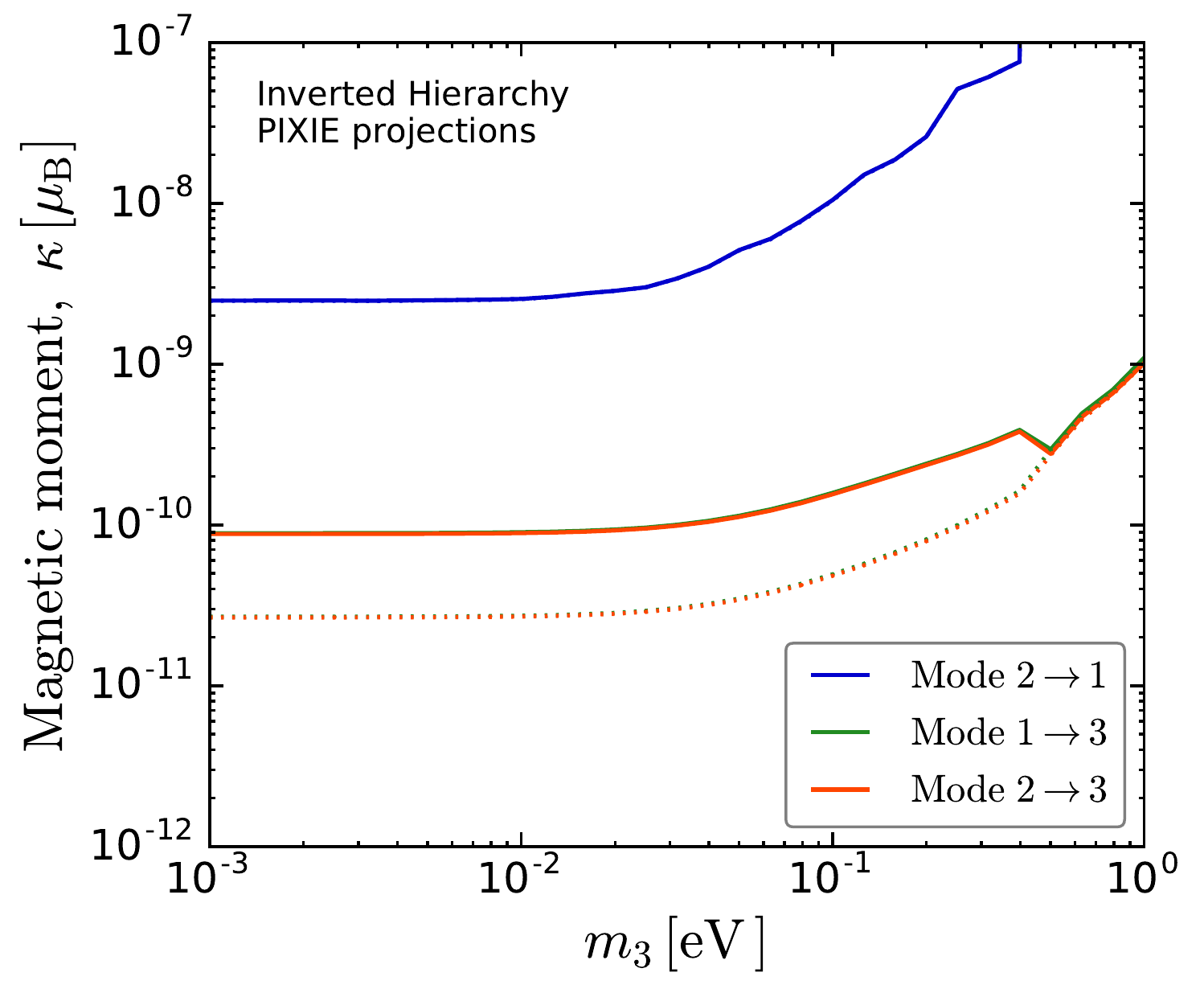}
\caption{\textbf{Projected 95\% C.L.~upper limits on the magnetic moment of neutrinos, as a function of the lightest neutrino mass for a PIXIE-like experiment.} The effective transition magnetic moment $\kappa_{ij}$ is related to the decay rate $\Gamma_{ij}$ by Eq.~\eqref{eq:kappa}. Dotted lines correspond to the upper limits assuming that the residual Galactic contamination $G_0$ is fixed to zero. Results shown are for NH (\textbf{left panel}) and IH (\textbf{right panel}).}
\label{fig:GroupD kappa}
\end{figure*}

We convert the projected upper limits on the neutrino lifetime
into limits on the neutrino transition magnetic and electric moments
$\kappa_{ij}$.
The result is shown in Fig.~\ref{fig:GroupD kappa}, where again dotted
lines show the case where the Galactic component is kept fixed.
The factor of $10^4$ improvement in neutrino lifetime constraints
translates to a factor of $10^2$ improvement in the magnetic moment
constraints.
For the lightest neutrino masses below 0.1~eV, the limit on $\kappa$
ranges from $10^{-8}\;\mu_B$ down to $3 \times 10^{-11}\;\mu_B$,
depending on the hierarchy and assumptions about Galactic
contamination.
In particular, we note that for the 23 mode, lightest neutrino mass
lower than $10^{-2}$~eV and minimal Galactic contamination, constraints
from a PIXIE-like experiment may be competitive with the best lab-based
$\nu$--$e$ scattering experiments (cf., constraints from BOREXINO give
$\mu_{\nu}<2.8\times 10^{-11} \mu_B$ \cite{Borexino2017} at 90\% C.L.).

\section{Conclusions}
\label{sec:conclusions}

In this work, we have presented updated constraints on the neutrino
radiative decay lifetime $\tau$ from FIRAS measurements of the CMB
intensity spectrum, introducing a number of refinements compared to
previous work~\cite{Mirizzi2007}.
We include spectral distortions from photon absorption by neutrinos (not
only from neutrino decays), as well as calculating decay and absorption
rates taking into account the momentum distribution of the cosmic
neutrino background.
In our analysis, we simultaneously fit the neutrino decay rate $\Gamma =
1/\tau$ along with other nuisance parameters, including the temperature
deviation $\Delta T$, $\mu$- and $y$-distortions and the residual
Galactic contamination $G_0$.
These lead to more accurate and robust limits than previously presented.

We find that the effects of absorption and decay are comparable for
neutrino masses $\mathcal{O}(0.1) \,\mathrm{eV}$ and larger. We also
find that the approximate formalism (assuming that the cosmic neutrinos
are at rest) may overestimate the spectral distortions by around
50\%. Finally, we find strong anti-correlation between the decay rate
$\Gamma$ and the other nuisance parameters in the analysis, weakening
the constraints on the neutrino lifetime unless a clear spectral feature
is produced in the FIRAS frequency range (as is that case where the
lightest mass lies in $[10^{-2}, 10^{-1}]\,\mathrm{eV}$ for the 12
mode).
While these effects should tend to weaken our constraints, we in
fact find stronger constraints than previous analyses
\cite{Mirizzi2007}, in some cases by around an order of magnitude,
although the source of this discrepancy is not clear.
In particular, we find $\tau_{12} \gtrsim 4 \times 10^{21}\,\mathrm{s}$
for the 12 decay mode in the normal hierarchy and $\tau_{12}
\gtrsim10^{22}\,\mathrm{s}$ in the inverted hierarchy.
For the 13 and 23 modes, there are no sharp spectral features in the
FIRAS frequency range, leading to weaker limits, $\tau_{13} \sim
\tau_{23} \gtrsim 10^{19} \,\mathrm{s}$.
The corresponding constraints on the neutrino magnetic moment lie in the
range $10^{-8}$--$10^{-7}\,\mu_B$.

We have also explored projected constraints from future precision CMB
spectral measurements, focusing on the proposed PIXIE
experiment~\cite{2011JCAP...07..025K}.
With an improvement in measurement sensitivity of around three orders of
magnitude compared to FIRAS, PIXIE should be able to constrain the
radiative decay lifetime of the neutrino at the level of $\tau \gtrsim
10^{23}$--$10^{25}\,\mathrm{s}$ depending on the neutrino mass and
hierarchy.
If residual Galactic contamination in the CMB spectrum is well
constrained, a PIXIE-like experiment may probe magnetic moments down to
$\kappa \lesssim 3 \times 10^{-11} \,\mu_B$ for the 13 and 23 modes.
While still one order of magnitude weaker than constraints from stellar
physics \cite{Raffelt1990,Raffelt:1999gv,Arceo-Diaz:2015pva}, such a
constraint would be competitive with current lab-based constraints from
$\nu$--$e$ scattering measurements \cite{Beda2013,Borexino:2017fbd}.
Further improvements in sensitivity, as proposed by the PRISM experiment
\cite{PRISMcollab}, would lead to still stronger bounds on the neutrino
lifetime and magnetic moment, making precision CMB spectral measurements
a competitive and complementary tool for probing New Physics in the
neutrino sector.

\begin{acknowledgments}
This work is supported partly by GRAPPA Institute at the University of
 Amsterdam (SA and BJK) and JSPS KAKENHI Grant Number JP17H04836 (SA).

This project has been carried out in the context of the ``ITFA 
 Workshop'' course, which is part of the joint bachelor programme in Physics and
 Astronomy of the University of Amsterdam and the Vrije Universiteit
 Amsterdam, for bachelor students (JLA, WMB, EB, JB, SB, GL, MR, DRvA,
 and HV).
The actual work was done in three independent groups A, C, and D (group
 B did not survive) during a four-week period of January 2018.
The group A (EB, GL, and HV) worked on theoretial calcluations of neutrino
 decay and absorption, having contributed to Secs.~\ref{sec:approximate
 formulae} and \ref{sec:theory results} and made
 Figs.~\ref{fig:absorp_12}--\ref{fig:ex_vs_approx_0.1}.
The group C (JLA, JB, and MR) worked on the COBE-FIRAS data analysis, and
 contributed to Sec.~\ref{sec:analysis} including
 Figs.~\ref{fig:tau}--\ref{fig:tau_constraints_NH_approx}.
The group D (WMB, SB, and DRvA) made future projection for PIXIE, having
 contributed to Sec.~\ref{sec:future} including
 Figs.~\ref{fig:GroupD tau} and \ref{fig:GroupD kappa}.
In addition, all the groups gave substantial contributions to
 Sec.~\ref{sec:introduction} by studying the relevant literature for
 each subject.
\end{acknowledgments}

\appendix

\section{Emissivity of photons}
\label{sec:appendix A}

In this section, we derive exact formulae for decay and absorption
intensities without making any assumptions.

\subsection{Decay}

From kinematics of the  decay $\nu_j \to \nu_i + \gamma$, one obtains
\begin{equation}
p_\gamma = p_{\gamma}^d(p_\nu,\mu) \equiv \frac{\Delta
m_{ij}^2}{2\left(\sqrt{p_\nu^2+m_j^2}-p_\nu\mu\right)}
\label{eq:kinematics decay}
\end{equation}
where $p_\nu =  |\bm{p}_\nu|$, $p_\gamma = |\bm{p}_\gamma|$,
$\bm{p}_\nu$ and $\bm{p}_\gamma$ are the momentum of $\nu_j$ and
$\gamma$, respectively, and $\mu = \bm{p}_\nu \cdot \bm{p}_\gamma  /
(p_\nu p_\gamma)$.
Alternatively, rewriting Eq.~(\ref{eq:kinematics decay}) for $\mu$
gives
\begin{equation}
 \mu = \mu_d(p_\gamma, p_\nu) \equiv
  \sqrt{1+\frac{m_j^2}{p_\nu^2}}-\frac{\Delta 
  m_{ij}^2}{2p_\nu p_\gamma}.
\label{eq:mu kinematics decay}
\end{equation}
In order for the decay to happen, the variables and parameters $(p_\nu,
p_\gamma, m_j, \Delta m_{ij}^2)$ have to satisfy $\mu_d^2 < 1$.
The momentum of the final-state neutrino $\nu_i$ is then obtained as
\begin{equation}
 p_{\nu_i}^d(p_\gamma,p_\nu) = \sqrt{p_\gamma^2+p_\nu^2-2p_\gamma
  p_\nu\mu_d(p_\gamma,p_\nu)}.
\end{equation}

The photon emissivity at energy $p_\gamma$ is obtained as a similar
equation as Eq.~(\ref{eq:emissivity decay}) but by replacing the
neutrino number density with the neutrino phase space density integrated
over momentum space:
\begin{eqnarray}
 P_{\rm dec}(p_\gamma) &=& \frac{g_{\nu}}{(2\pi)^2}\int
  \frac{d^3p_\nu}{e^{p_\nu/T_\nu}+1}\frac{m_j
  \Gamma}{\sqrt{p_\nu^2+m_j^2}}
  \nonumber\\&&\times
  p_\gamma
  \delta\left(p_\gamma - p_\gamma^d(p_\nu,\mu)\right)
  (1+f_{\rm CMB}(p_\gamma))
  \nonumber \\ &&\times
  \left[1-\frac{1}{e^{p_{\nu_i}^d(p_\gamma,p_\nu)/T_\nu}+1}\right],
  \nonumber\\
 &=& \frac{\Gamma m_j \Delta m_{\ij}^2}{4\pi^2} \frac{1+f_{\rm
  CMB}(p_\gamma)}{p_\gamma}
  \nonumber\\&&\times
  \int_0^\infty 
  \frac{dp_\nu p_\nu}{(e^{p_\nu/T_\nu}+1)\sqrt{p_\nu^2+m_j^2}}
  \nonumber\\&&\times
  \left[1-\frac{1}{e^{p_{\nu_i}^d(p_\gamma,p_\nu)/T_\nu}+1}\right]
  \nonumber\\&&\times 
  \int_{-1}^1 d\mu \delta\left(\mu - \mu_d(p_\gamma,p_\nu)\right),
\end{eqnarray}
where $g_\nu = 2$ is the number of helicities of the neutrino, the
lifetime of $\nu_j$ is longer than its proper lifetime $\tau$ by a
Lorentz factor of $(p_\nu^2+m_j^2)^{1/2}/m_j$, the stimulated emission
is taken into account with the term $1+f_{\rm CMB}(p_\gamma)$, and the
term in the square bracket represents the Pauli blocking for the final
neutrino state with momentum $p_{\nu_i}^d$.
In the second equality, we changed the $\delta$-function of $p_\gamma$
to that of $\mu$, by using $\delta\left(p_\gamma-p_\gamma^d\right) =
\left|{\partial p_\gamma^d}/{\partial \mu}\right|^{-1}\delta\left(\mu-\mu_d\right)$,
and $|\partial p_\gamma^d / \partial \mu| = 2p_\gamma^2 p_\nu / \Delta
m_{ij}^2$.

The $\mu$-integral over its $\delta$-function gives nonzero value (i.e.,
1) only if $\mu_d^2 < 1$.
By rearranging Eq.~(\ref{eq:mu kinematics decay}), we find the condition
to be equivalent to $x > |t-t^{-1}|y/2$, where $x = p_\nu/T_\nu$, $y
= m_j/T_\nu$, and $t = 2p_\gamma m_j/\Delta m_{ij}^2$.
We also note $\mu_d = \left[(x^2+y^2)^{1/2}-y/t\right]/x$.

Finally, we evaluate the emissivity at redshift $z$ and observed energy
of $\epsilon_\gamma$.
We can use all the equations derived thus far with replacements:
$p_\gamma \to (1+z)\epsilon_\gamma$, $p_\nu \to (1+z)p_\nu$, and $T_\nu
\to (1+z)T_\nu$.
The emissivity of the neutrino decay is then obtained as
Eq.~(\ref{eq:P decay exact}).

\subsection{Absorption}

For the absorption $\nu_i + \gamma \to \nu_j$, the kinematics relations
are
\begin{eqnarray}
 p_\gamma &=& p_\gamma^a(p_\nu,\mu) \equiv \frac{\Delta
  m_{ij}^2}{2\left(\sqrt{p_\nu^2+m_i^2} - p_\nu\mu\right)},
  \\
 \mu &=& \mu_a(p_\gamma,p_\nu) \equiv \sqrt{1+\frac{m_i^2}{p_\nu^2}} -
  \frac{\Delta m_{ij}^2}{2p_\nu p_\gamma},
\end{eqnarray}
where $p_\nu$ is (norm of) the momentum of $\nu_i$; definitions of the
other quantities are the same as the case of decay.
The momentum of the final-state neutrino $\nu_j$ is then
\begin{equation}
 p_{\nu_j}^a(p_\gamma,p_\nu) = \sqrt{p_\gamma^2+p_\nu^2 + 2p_\gamma
  p_\nu \mu_a(p_\gamma,p_\nu)}.
\end{equation}

The center-of-mass energy of the initial state $E$ is given by $E^2 = m_i^2 +
2p_\gamma \left[(p_\nu^2+m_i^2)^{1/2} - p_\nu\mu\right]$, and
the $\delta$-function of $E$ can be replaced with that of $\mu$ through
$\delta (E-m_j) = |\partial E / \partial \mu|^{-1}\delta (\mu-\mu_a)$
with $|\partial E / \partial \mu| = p_\gamma p_\nu / m_j$.
The absorption cross section [Eq.~(\ref{eq:absorption cross section})]
then becomes
\begin{equation}
 \sigma (p_\gamma,p_\nu,\mu) = \frac{4\pi^2 m_j^3 \Gamma}{(\Delta
  m_{ij}^2)^2p_\gamma p_\nu} \delta (\mu - \mu_a).
  \label{eq:sigma abs of mu}
\end{equation}

The absorption emissivity is given as product of the phase space
densities of both $\gamma$ and $\nu_i$, multiplied by the absorption
cross section as follows:
\begin{eqnarray}
 P_{\rm abs}(p_\gamma)dp_\gamma &=& -p_\gamma
  \frac{g_\gamma}{(2\pi)^3}f_{\rm CMB}(p_\gamma)d^3p_\gamma
  \nonumber\\&&{}\times
  \frac{g_\nu}{(2\pi)^3}\int
  \frac{d^3p_\nu}{e^{p_\nu/T_\nu}+1}\sigma(p_\gamma,p_\nu,\mu)
  \nonumber\\&&{}\times
  \left[1-\frac{1}{e^{p_{\nu_j}^a(p_\gamma,p_\nu)/T_\nu}+1}\right],
\end{eqnarray}
where $g_\gamma = 2$ is the number of polarization states of the photon.
Using Eq.~(\ref{eq:sigma abs of mu}) and performing the $\mu$-integral
over the $\delta$-function that yields a nonzero value only when
$\mu_a^2 < 1$, one obtains
\begin{eqnarray}
 P_{\rm abs}(p_\gamma) &=& -\frac{2}{\pi^2}\frac{m_j^3\Gamma}{(\Delta
  m_{ij}^2)^2}p_\gamma^2 f_{\rm CMB}(p_\gamma)
  \nonumber\\&&{}\times
  \int_0^\infty \frac{dp_\nu p_\nu}{e^{p_\nu/T_\nu}+1}
  \Theta\left(1-\mu_a^2\right(p_\gamma,p_\nu))
  \nonumber\\&&{}\times
  \left[1-\frac{1}{e^{p_{\nu_j}^a(p_\gamma,p_\nu)/T_\nu}+1}\right],
\end{eqnarray}
where $\Theta$ is the Heaviside step function.
As in the case of decay, this constraint, $\mu_a^2<1$, is equivalent to
$x > |t-t^{-1}|y/2$, with $x =  p_\nu/T_\nu$, $y = m_i / T_\nu$, and $t
= 2p_\gamma m_i / \Delta m_{ij}^2$.

Lastly, with replacements $p_\gamma \to (1+z)\epsilon_\gamma$, $p_\nu
\to (1+z)p_\nu$, $T_{\rm CMB} \to (1+z)T_{\rm CMB}$, and $T_\nu \to
(1+z)T_\nu$, we arrive at Eq.~(\ref{eq:P abs exact}).

\bibliographystyle{apsrev4-1}
\bibliography{refs}

\end{document}